\documentclass{aa}
\usepackage{graphicx}
\usepackage{setspace}

\begin{document}

\newcommand{\fe}{[\ion{Fe}{ii}]}
\newcommand{\s}{[\ion{S}{ii}]}
\newcommand{\h}{H$_2$}
\newcommand{\kms}{km\,s$^{-1}$}
\newcommand{\um}{$\mu$m}
\newcommand{\lam}{$\lambda$}

\title{A combined optical/infrared spectral diagnostic analysis of the
HH1 jet
\thanks{Based on observations collected at the European Southern 
Observatory, La Silla, Chile (ESO programmes 070.C-0396(A), 070.C.-0396(B))}}
\author{Brunella Nisini\inst{1} \and Francesca Bacciotti\inst{2} \and Teresa  Giannini\inst{1} \and
 Fabrizio Massi\inst{2} \and Jochen Eisl\"offel\inst{3} \and Linda  Podio\inst{2,4} \and Thomas P. Ray\inst{5}}
\offprints{Brunella Nisini, nisini@mporzio.astro.it}

\institute{INAF-Osservatorio Astronomico di Roma, Via di Frascati 33, I-00040 Monteporzio Catone,
Italy \and INAF-Osservatorio Astrofisico di Arcetri, Largo E. Fermi 5, I-50125 Florence, Italy
\and Th\"uringer Landessternwarte Tautenburg, Sternwarte 5, D-07778 Tautenburg, Germany
\and Dipartimento di Astronomia e Scienza dello Spazio, Universit\'a degli Studi 
di Firenze, Largo E. Fermi 2, I-50125 Firenze, Italy
\and School of Cosmic Physics, Dublin Institute for Advanced Studies, 5 Merrion Square, Dublin 2, Ireland}
 
%
\date{Received date; Accepted date}
%
%
%
\titlerunning{Combined optical/infrared diagnostics of the HH1 jet}
\authorrunning{}

\abstract{
Complete flux-calibrated spectra covering the spectral range 
from 6000\AA\, to 2.5\um\, have been obtained along the HH1 jet and analysed 
in order to explore the potential of a combined optical/near-IR
diagnostic applied to jets from young stellar objects.
The main physical parameters (visual extinction, electron temperature and
density, ionization fraction and total density) have been derived 
along the jet using various diagnostic line ratios.
This multi-line analysis shows, in each spatially unresolved knot,
 the presence of zones at different excitation conditions, as expected 
 from the cooling layers behind a shock front.
In particular, a density stratification in the jet is evident
from ratios of various lines of different critical density. We measure 
electron densities in the range 6\,10$^{2}$-3\,10$^{3}$ cm$^{-3}$ with the 
\s\, optical doublet lines, 4\,10$^{3}$-10$^{4}$ cm$^{-3}$ with the
near-IR \fe\, lines, and 10$^{5}$-10$^{6}$ cm$^{-3}$ with
optical \fe\, and CaII lines. The electron temperature also shows variations,
with values  between 8000-11000 K derived from optical/near-IR \fe\, lines
and 11000-20000 K from a combined diagnostic employing
 optical [\ion{O}{i}] and [\ion{N}{ii}] lines. Thus \fe\, lines originate in
a cooling layer located at larger distances from the shock front than that
generating the optical lines, where the compression is higher and the
temperature is declining.

The derived parameters were used to measure the mass flux along the
jet, adopting different procedures, the advantages and
limitations of which are discussed. The \fe\,1.64\um\, line
luminosity turns out to be more suitable to measure $\dot{M}_{jet}$ 
than the optical lines, since it samples a fraction of 
the total mass flowing through a knot larger than the
[\ion{O}{i}] or \s\, lines.
$\dot{M}_{jet}$ is high in the initial part of the flow 
($\sim$2.2\,10$^{-7}$ M$_{\odot}$\,yr$^{-1}$) but decreases
by about an order of magnitude further out. Conversely, the mass flux
associated with the warm molecular material is low,
$\dot{M}_{H2}$$\sim$10$^{-9}$ M$_{\odot}$\,yr$^{-1}$, and does not 
show appreciable variations along the jet.
We suggest that part of the mass flux in the external regions 
is not revealed in optical and IR lines because it is 
associated with a colder atomic component, which may be
traced by the far-IR [\ion{O}{i}]63\um\, line.
  
Finally, we find that the gas-phase abundance of refractory species, such as Fe, C, Ca, 
and Ni, is lower than the solar value, with the lowest values (between 10
and 30\% of solar) derived in the inner and densest regions. This 
suggests a
significant fraction of dust grains may {\it still be present} in the jet beam,  
imposing constraints on the efficiency of grain destruction by multiple low-velocity
shock events.

}
\maketitle
%

\section{Introduction}

Highly-collimated jets of matter from young stellar objects (YSOs) are a common phenomena 
in star formation. These flows exhibit rich emission-line spectra 
over a wide range of wavelengths, from the UV to the radio. 
Many of the important tracers of physical conditions in the jet lie in the 
optical and near IR (NIR).
In the optical, prominent lines such as [\ion{O}{i}]\lam6300, 
H$\alpha$, [\ion{N}{ii}]$\lambda$6583, \s$\lambda\lambda$6716,6731, have
been widely used to derive electron density, temperature, and ionization
fraction along the jet beams (e.g. Bacciotti \& Eisl\"offel 1999, hereafter BE99, 
Hartigan et al. 1994). The NIR spectrum, on the other hand, comprises both
important atomic lines such as \fe, Pa$\beta$, \s, and [\ion{N}{i}], giving 
complementary information
on the excitation conditions (Nisini et al. 2002), as well as intense 
H$_2$ transitions 
from various vibrational levels, tracing the colder molecular part of the
post-shocked gas (e.g. Giannini et al. 2004, Eisl\"offel et al. 2000).
So far, the optical and NIR spectra of the same object have never been  
compared through combined optical/NIR diagnostics. 
Moreover, studies in both the optical and  
the infrared have mostly concentrated on limited spectral ranges, 
e.g. around 6500\AA\, and 2\um. Hence the diagnostic capabilities of
lines in other parts of the spectrum have not been
fully exploited.
Only recently has the potential of transitions from 0.7 to
1.5\um\, been highlighted by Giannini et al. (2002) and Hartigan et al. (2004).

The aim of the work described in this paper is to fill this gap and build 
a set of tools for a combined optical/NIR analysis of spectra of stellar jets.
The advantage of such a multi-wavelength approach is twofold. 
On the one hand it allows us to employ
ratios between lines from the same species that are well separated in wavelength 
thereby providing more stringent constraints on excitation conditions. 
This is the case, for example,  for the \fe8620\AA/1.64\um\, and the \s6716,6731\AA/1.03\um\,
ratios, which are very sensitive to the gas electron temperature (Nisini et al. 
2002, Pesenti et al. 2002). On the other hand, such 
an approach gives us the possibility of probing the different components of the jet 
cooling layers,
where strong gradients are expected along the jet axis, that are not spatially 
resolved with current instrumentation. 
This, in turn, allows a better determination of 
parameters, such as the mass flux in the jet, which are fundamental to
its dynamics and other properties.
A correct derivation of this quantity is particularly important 
in heavily embedded sources, where the mass flux is the primary indirect 
manifestation of protostellar accretion activity.

In addition, a wide wavelength coverage gives us the possibiliy of
observing lines of less abundant species that nevertheless contain
important 
diagnostic information. In particular, lines from many refractory species
(Ca, Ni, Cr in addition to Fe and C) can be used to derive the amount
of gas depletion in the jet, thus setting constraints to the
degree of dust grain destruction by shocks. The degree of grain 
destruction in shocks, and the relative gas-phase abundances among different 
species, depend on the shock properties, but also on the grain structure
 (Jones 2000). The latter, however, is not completely understood,
although the grain composition is of crucial importance in many processes, 
spanning from the physics of molecular clouds to the formation of planets.

For our purpose we have observed the HH1 jet by obtaining flux calibrated spectra 
covering the wavelength range from 0.65 to 2.5\um, as described in Section 2.
This jet, emanating from the radio source 
VLA1 (Strom et al. 1985) and located in the Orion L1641 molecular (D=460 pc), is
well suited to test combined optical/IR analysis. It powers the prototype HH 
objects HH1/2 and is bright both
in optical (Eisl\"offel et al. 1994, Hester et al. 1998, 
Reipurth et al. 2000) NIR H$_{2}$ and \fe\, lines (Davis et al. 1994, Noriega-Crespo et al. 1994, Davis et al. 2000). 
A limited set of optical forbidden, as well as H$_{2}$, emission lines were
analysed in the past by various authors (Solf \& B\"ohm 1991, Noriega-Crespo et al. 1997, 
Eisl\"offel et al. 2000, Medves 2003).
Here we use our new combined optical/near-IR observations to derive the relevant
physical parameters ($A_V$, $n_e$, $T_e$, $x_e$, and $n_{\rm H}$) and their
variation along the HH1 jet, using different line ratios capable
of diagnosing regions with different excitation conditions. This is described in Section 3,
while the diagnostic capabilities of specific lines are illustrated in some 
detail in Appendix A. In Appendix B we discuss the limitations 
of the widely used \fe\,1.64/1.25\um\, ratio to measure 
extinction along the line of sight. 

In Section 4 we estimate the jet mass flux rate by adopting different 
lines and methods, discussing the limitations and advantages of the
different procedures (Section 4).
Finally, in Section 5, we will set constraints on
the efficiency of shocks to disrupt dust grains by deriving the gas-phase
abundance of different refractory species.
In Section 6 we summarize our results discussing their implications 
for the understanding of HH jet properties.

\section{Observations and results}

\subsection{ESO 3.6m-EFOSC2 and NTT-SofI observations}

Since the aim of this project is to apply a combined optical/NIR line
analysis, we obtained spectra that are as homogeneous as possible in the two
wavelength ranges. Observations with the ESO 3.6-m and NTT telescopes on 
La Silla, Chile, were carried out during adjacent nights (7-8 January
and 11-12 January 2003 for the optical and IR observations, 
respectively). The instruments,
EFOSC2 for the optical and SofI for the infrared, had the
same slit width (1$\arcsec$), corresponding to a similar spectral resolution
between 500 and 600. The spatial scale of the two cameras are also comparable,
being 0$\farcs$314/pixel for EFOSC2 and 0$\farcs$29/pixel for SofI. 
The slits were aligned along the HH1 jet at a P.A. of -36$^{o}$.
The EFOSC2 observations covered the spectral 
range from 6015 to 10320\AA\, in a single grism setting, while the 
NTT observations were obtained with the blue (0.95-1.64\um)
and red (1.53-2.52\um) grisms. Integration times were 1800s for 
EFOSC2 and 2400s for each of the two SofI grisms.
Data reduction was performed using standard IRAF tasks. In both 
the optical and IR spectra we corrected for the atmospheric spectral response 
by dividing the object spectra by the spectrum of an O8 star 
showing telluric lines. The spectra were wavelength calibrated with 
a xenon-argon lamp, giving an uncertainty smaller than the spectral 
resolution element ($\sim$ 20\AA). Flux calibrations were performed 
using photometric standards observed at an air mass difference not 
larger than 0.3 with respect to our target observations.

\subsection{Extraction of the spectra and intercalibration}

\begin{figure}[!ht]
\resizebox{\hsize}{!}{\includegraphics{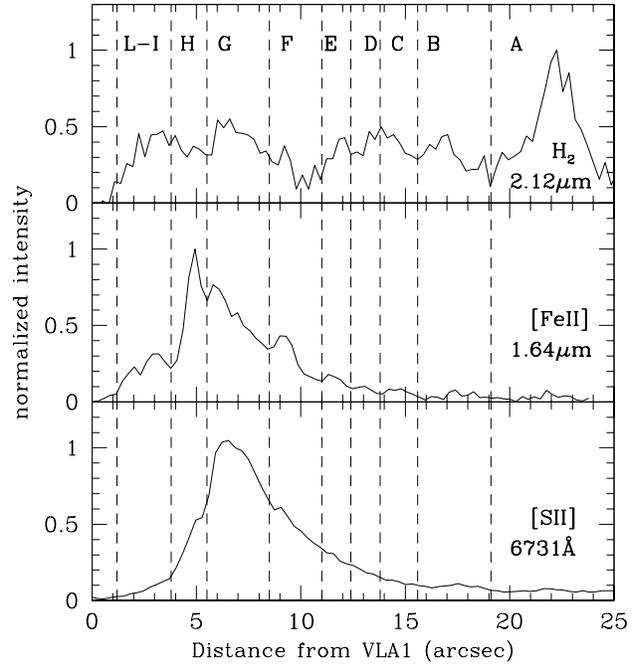}}
\caption{\label{fig:line_em}
Profiles of the \s, \fe\, and \h\, emission along the HH1 jet. 
Vertical dashed lines indicate the range coadded to obtain individual
knot fluxes. The \fe\, 
profile was used to define the knot separation.}
\end{figure}

At the end of the reduction procedure, we had three individually calibrated
 spectral images
that have to be combined to produce single spectra of the same 
regions. The spatial zero point was set 
to the position of the jet source, VLA1. To 
determine this in our frames, we used the bright Cohen-Schwartz (CS) 
star, adopting for it an angular separation of 35$\farcs$9 from VLA1 
(Rodr\'{\i}guez et al. 2000). 
Figure \ref{fig:line_em} reports the emission profiles of bright lines observed
in the three different images, namely \s\,6731\AA, \fe\,1.64\um\, and \h\,2.12 $\mu$m,
that trace different exitation conditions and/or extinction.
From the \fe\, profiles, nine contiguous emission knots have been 
identified, named A to I following the nomenclature of Eisl\"offel et al. (1994)
and Bally et al. (2002) 
\footnote{Bally et al. (2002) named the knots in the jet from A$_J$ to I$_J$
in order of decreasing distance from VLA1.
They also found two additional knots, A and B, close 
to knot G$_J$ but at a different 
position angle, and thus probably belonging to a different jet.}.
Their spatial extent is reported in Table 2. 
We then extracted the spectra of the individual knots and computed 
the line fluxes by fitting the profiles with single or, where blending
is present, double 
Gaussian profiles. Errors to these line fluxes are measured from the rms of the
adjacent continuum.

To intercalibrate fluxes for the three spectral segments of the same 
knot, we used the spectrum from the SofI-blue grism as a reference,  normalizing 
the EFOSC2 spectrum to it with the [\ion{C}{ii}]\lam\lam9824,9850 lines,
and the SofI-red grism spectrum with the \fe\,1.64$\mu$m line. We 
note that the intercalibration factors may differ from unity 
and change from knot to knot, for a variety of causes. 
These include different atmospheric conditions during the
observations, a small misalignment of the optical and IR slits, and 
slighty different spatial sampling of the instruments. The net result were 
intercalibration 
factors ranging between 0.6 and 1.1 for EFOSC2 and the SofI blue 
grism, and between 0.9 and 1.5 for the blue and red SofI 
grisms. For those knots where the intercalibration lines were not observed, 
average factors, derived for the other knots, were used.

\begin{figure*}
\resizebox{\hsize}{!}{\includegraphics{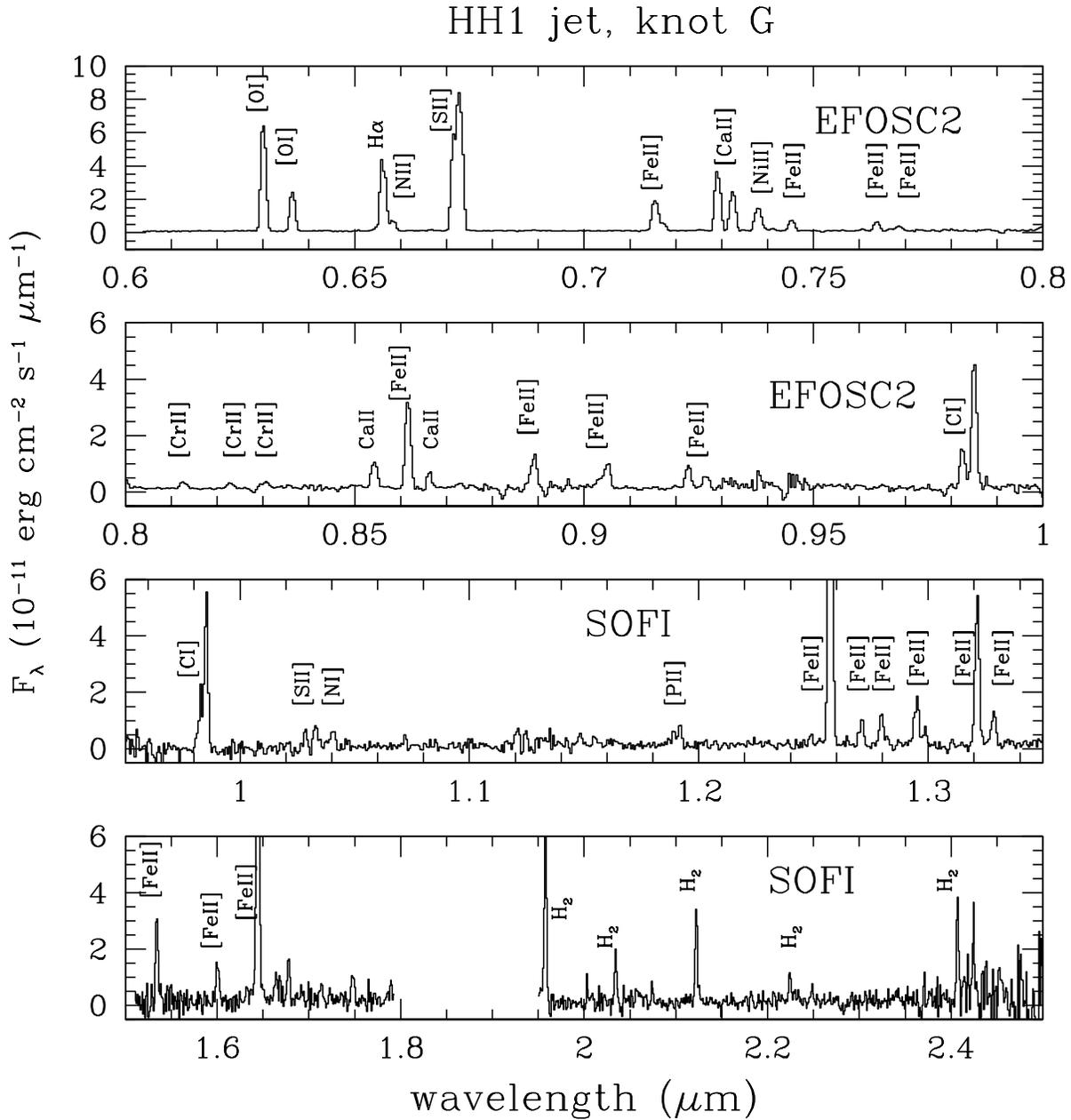}}
   \caption{\label{fig:gspectrum} Combined EFOSC2 + SofI spectrum from 0.6
   to 2.5$\mu$m of knot G of the HH1 jet. The stronger lines are identified.}
\end{figure*}

Figure \ref{fig:gspectrum} shows the complete, intercalibrated spectrum of 
knot G -- the brightest in the optical --and in Table 1 we report 
fluxes and identification for all the detected lines in this knot. 
These, as a whole, indicate low excitation conditions, as expected for HH objects. 
In addition, H$_2$ lines from the 1-0 and 2-1 vibrational series were also 
observed.


\begin{table*}
\caption[]{Lines observed in knot G of the HH1 jet }
\vspace{0.5cm}
\begin{tabular}{cccc}
\hline
 Line id. &  $\lambda^{a}$ & $F$ & $\Delta~F$\\
          &    & \multicolumn{2}{c}{erg\,cm$^{-2}$\,s$^{-1}$} \\[+5pt]  
\hline\\[-5pt] 
~[\ion{O}{i}]~$^1\!D_{2}-^3\!P_{2}$ & 6300  &  1.24\,10$^{-14}$ & 2.46\,10$^{-17}$\\
~[\ion{O}{i}]~$^1\!D_{2}-^3\!P_{1}$ &  6364  &  4.54\,10$^{-15}$   &   2.78\,10$^{-17}$\\
~H$\alpha$ &                          6562  &  9.34\,10$^{-15}$   &   3.93\,10$^{-17}$\\
~[\ion{N}{ii}]~$^1\!D_{2}-^3\!P_{2}$ &  6583  &  1.43\,10$^{-15}$  &  4.38\,10$^{-17}$\\
~[\ion{S}{ii}]~$^2\!D_{5/2}-^4\!S_{3/2}$ &  6716  & 2.86\,10$^{-14}$   &  5.83\,10$^{-17}$\\
~[\ion{S}{ii}]~$^2\!D_{3/2}-^4\!S_{3/2}$ &  6731 &   &  \\
~[\ion{Fe}{ii}]~$a^2\!G_{9/2}-a^4\!F_{9/2}$ &  7155  &  4.02\,10$^{-15}$   &  2.99\,10$^{-17}$\\
~[\ion{Fe}{ii}]~$a^2\!G_{7/2}-a^4\!F_{7/2}$ &  7172  &  1.34\,10$^{-15}$   &  3.56\,10$^{-17}$\\
~[\ion{Ca}{ii}]~$^2\!D_{5/2}-^2\!S_{1/2}$ &  7291  &  7.52\,10$^{-15}$   &   3.22\,10$^{-17}$\\
~[\ion{Ca}{ii}]~$^2\!D_{3/2}-^2\!S_{1/2}$ &  7324  &  4.91\,10$^{-15}$   &   3.18\,10$^{-17}$\\
~[\ion{Ni}{ii}]~$^2\!F_{7/2}-^2\!D_{5/2}$ &  7378  &  3.10\,10$^{-15}$   &   3.50\,10$^{-17}$\\
~[\ion{Ni}{ii}]~$^2\!F_{5/2}-^2\!D_{3/2}$ &  7412 &   1.87\,10$^{-16}$   &  2.39\,10$^{-17}$\\
~[\ion{Fe}{ii}]~$a^2\!G_{9/2}-a^4\!F_{7/2}$ &  7453  & 1.25\,10$^{-15}$  &    2.31\,10$^{-17}$\\
~[\ion{Fe}{ii}]~$a^4\!P_{5/2}-a^6\!D_{7/2}$ &  7638 &  1.15\,10$^{-15}$   &   7.01\,10$^{-17}$\\
~[\ion{Fe}{ii}]~$a^4\!P_{1/2}-a^6\!D_{3/2}$ &  7665 &   2.85\,10$^{-16}$   &   7.10\,10$^{-17}$\\
~[\ion{Fe}{ii}]~$a^4\!P_{3/2}-a^6\!D_{5/2}$ &  7687  &  8.53\,10$^{-16}$   &   1.29\,10$^{-16}$\\
~[\ion{Cr}{ii}]~$^6\!D_{9/2}-^6\!S_{5/2}$ &  8000  &  5.90\,10$^{-16}$   &   5.71\,10$^{-17}$\\
~[\ion{Cr}{ii}]~$^6\!D_{7/2}-^6\!S_{5/2}$ &  8125 &   4.88\,10$^{-16}$  &    3.45\,10$^{-17}$\\
~[\ion{Cr}{ii}]~$^6\!D_{5/2}-^6\!S_{5/2}$ &  8229 &   4.40\,10$^{-16}$  &    7.58\,10$^{-17}$\\
~[\ion{Cr}{ii}]~$^6\!D_{3/2}-^6\!S_{5/2}$ &  8308 &   6.03\,10$^{-16}$  &    1.46\,10$^{-16}$\\
~\ion{Ca}{ii}~$^2\!P^o_{3/2}-^2\!D_{5/2}$ &  8542 &   1.96\,10$^{-15}$  &    9.78\,10$^{-17}$\\
~[\ion{Fe}{ii}]~$a^4\!P_{5/2}-a^4\!F_{9/2}$ &  8617 &  5.98\,10$^{-15}$   &   7.32\,10$^{-17}$\\
~\ion{Ca}{ii}~$^2\!P^o_{1/2}-^2\!D_{3/2}$ &   8662 &  1.11\,10$^{-15}$   &   1.44\,10$^{-16}$\\
~[\ion{Fe}{ii}]~$a^4\!P_{3/2}-a^4\!F_{7/2}$ &  8892 &  2.92\,10$^{-15}$  &    2.79\,10$^{-16}$\\
~[\ion{Fe}{ii}]~$a^4\!P_{1/2}-a^4\!F_{5/2} $ &  9034 & 7.83\,10$^{-16}$  &    1.58\,10$^{-16}$\\
~[\ion{Fe}{ii}]~$a^4\!P_{5/2}-a^4\!F_{7/2}$ &  9052 &  2.10\,10$^{-15}$   &   1.82\,10$^{-16}$\\
~[\ion{Fe}{ii}]~$a^4\!P_{3/2}-a^4\!F_{5/2}$ &  9227 &  1.51\,10$^{-15}$  &    1.52\,10$^{-16}$\\
~[\ion{Fe}{ii}]~$a^4\!P_{1/2}-a^4\!F_{3/2}$ &   9268 &   7.75\,10$^{-16}$  & 1.63\,10$^{-16}$\\
~[\ion{C}{i}]~$^1\!D_{2}-^3\!P_{1}$ &  9824  &  3.60\,10$^{-15}$   &  2.19\,10$^{-16}$\\
~[\ion{C}{i}]~$^1\!D_{2}-^3\!P_{2}$ &  9850  &  9.48\,10$^{-15}$   &  2.01\,10$^{-16}$\\
~[\ion{S}{ii}]\,$^2\!P_{3/2}-^2\!D_{3/2}$ &   1.029  &  8.77\,10$^{-16}$  &  2.10\,10$^{-16}$\\

\hline\\[-5pt]
\end{tabular}
\begin{tabular}{cccc}
\hline
 Line id. &  $\lambda^{a}$ & $F$ & $\Delta~F$\\
          &    & \multicolumn{2}{c}{erg\,cm$^{-2}$\,s$^{-1}$} \\[+5pt]  
\hline\\[-5pt] 

~[\ion{S}{ii}]\,$^2\!P_{3/2}-^2\!D_{5/2}$ &   1.032  &   1.86\,10$^{-15}$  &    2.82\,10$^{-16}$\\
~[\ion{S}{ii}]\,$^2\!P_{1/2}-^2\!D_{3/2}$ &   1.034  & 	  &  \\
~[\ion{N}{i}]\,$^2\!P_{3/2,1/2}-^2\!D_{5/2}$ &   1.040  &  1.77\,10$^{-15}$  &    4.35\,10$^{-16}$\\
~[\ion{N}{i}]\,$^2\!P_{3/2,1/2}-^2\!D_{3/2}$ &   1.041 &    &  \\
~[\ion{P}{ii}]~$^3\!P_{1}-^1\!D_{2}$  &  1.148 &   1.02\,10$^{-15}$  &    2.01\,10$^{-16}$\\
~[\ion{P}{ii}]~$^3\!P_{2}-^1\!D_{2}$  &  1.189 &   7.04\,10$^{-16}$  &    8.04\,10$^{-17}$\\
~[\ion{Fe}{ii}]~$a^4\!D_{5/2}-a^6\!D_{9/2}$ &   1.192 &   1.21\,10$^{-15}$  &    1.06\,10$^{-16}$\\
~[\ion{Fe}{ii}]~$a^4\!D_{3/2}-a^6\!D_{5/2}$ &   1.248 &   7.55\,10$^{-16}$  &    1.66\,10$^{-16}$\\
~[\ion{Fe}{ii}]~$a^4\!D_{5/2}-a^6\!D_{7/2}$ &   1.249     &   &  \\
~[\ion{Fe}{ii}]~$a^4\!D_{7/2}-a^6\!D_{9/2}$ &   1.258 &   2.90\,10$^{-14}$  &    1.62\,10$^{-16}$\\
~[\ion{Fe}{ii}]~$a^4\!D_{1/2}-a^6\!D_{1/2}$ &   1.271  &  2.06\,10$^{-15}$  &    2.32\,10$^{-16}$\\
~[\ion{Fe}{ii}]~$a^4\!D_{3/2}-a^6\!D_{3/2}$ &   1.280 &   2.57\,10$^{-15}$  &  2.45\,10$^{-16}$\\
~[\ion{Fe}{ii}]~$a^4\!D_{5/2}-a^6\!D_{5/2}$ &   1.295 &   4.10\,10$^{-15}$  &    2.23\,10$^{-16}$\\
~[\ion{Fe}{ii}]~$a^4\!D_{3/2}-a^6\!D_{1/2}$ &   1.299 &   1.22\,10$^{-15}$  &    1.52\,10$^{-16}$\\
~[\ion{Fe}{ii}]~$a^4\!D_{7/2}-a^6\!D_{7/2}$ &   1.321 &   1.03\,10$^{-14}$  &    1.77\,10$^{-16}$\\
~[\ion{Fe}{ii}]~$a^4\!D_{5/2}-a^6\!D_{3/2}$ &   1.329 &   2.22\,10$^{-15}$  &    1.72\,10$^{-16}$\\
~[\ion{Fe}{ii}]~$a^4\!D_{5/2}-a^4\!F_{9/2}$ &   1.534  &  4.39\,10$^{-15}$  &    1.19\,10$^{-16}$\\
~[\ion{Fe}{ii}]~$a^4\!D_{3/2}-a^4\!F_{7/2}$ &   1.601 &   3.15\,10$^{-15}$  &    2.31\,10$^{-16}$\\
~[\ion{Fe}{ii}]~$a^4\!D_{7/2}-a^4\!F_{9/2}$ &   1.645 &   3.27\,10$^{-14}$  &    1.92\,10$^{-16}$\\
~[\ion{Fe}{ii}]~$a^4\!D_{1/2}-a^4\!F_{5/2}$ &   1.665 &   1.64\,10$^{-15}$  &    2.38\,10$^{-16}$\\
~[\ion{Fe}{ii}]~$a^4\!D_{5/2}-a^4\!F_{7/2}$ &   1.678 &   3.42\,10$^{-15}$  &   2.99\,10$^{-16}$\\
~H$_2$~1--0 S(7)  &  1.748  &  1.52\,10$^{-15}$   &   3.50\,10$^{-16}$\\
~H$_2$~1--0 S(6)  &  1.788  &  1.95\,10$^{-15}$  &    3.62\,10$^{-16}$\\
~H$_2$~1--0 S(3)  &  1.958  &  1.77\,10$^{-14}$  &    2.45\,10$^{-16}$\\
~H$_2$~1--0 S(2)  &  2.034  &  1.99\,10$^{-15}$  &    1.98\,10$^{-16}$\\
~H$_2$~2--1 S(3)  &  2.074  &  8.99\,10$^{-16}$  &    9.02\,10$^{-17}$\\
~H$_2$~1--0 S(1)  &  2.122  &  5.27\,10$^{-15}$  &    1.93\,10$^{-16}$\\
~H$_2$~1--0 S(0)  &  2.224  &  1.57\,10$^{-15}$  &    1.66\,10$^{-16}$\\
~H$_2$~2--1 S(1)  &  2.249  &  8.89\,10$^{-16}$  &    1.98\,10$^{-16}$\\
~H$_2$~1--0 Q(1)  &  2.407  &  5.44\,10$^{-15}$  &    2.59\,10$^{-16}$\\
~H$_2$~1--0 Q(2)  &  2.414  &  1.60\,10$^{-15}$  &    1.84\,10$^{-16}$\\
~H$_2$~1--0 Q(3)  &  2.424  &  4.73\,10$^{-15}$  &    2.63\,10$^{-16}$\\
\hline\\[-5pt]
\end{tabular}

~$a$ Air wavelengths in \AA\, below 1\um\, and vacuum wavelengths in micron above. 
\end{table*}

\subsection{Origin of the observed lines}

\begin{figure*}[!ht]
\sidecaption
\includegraphics[width=12cm]{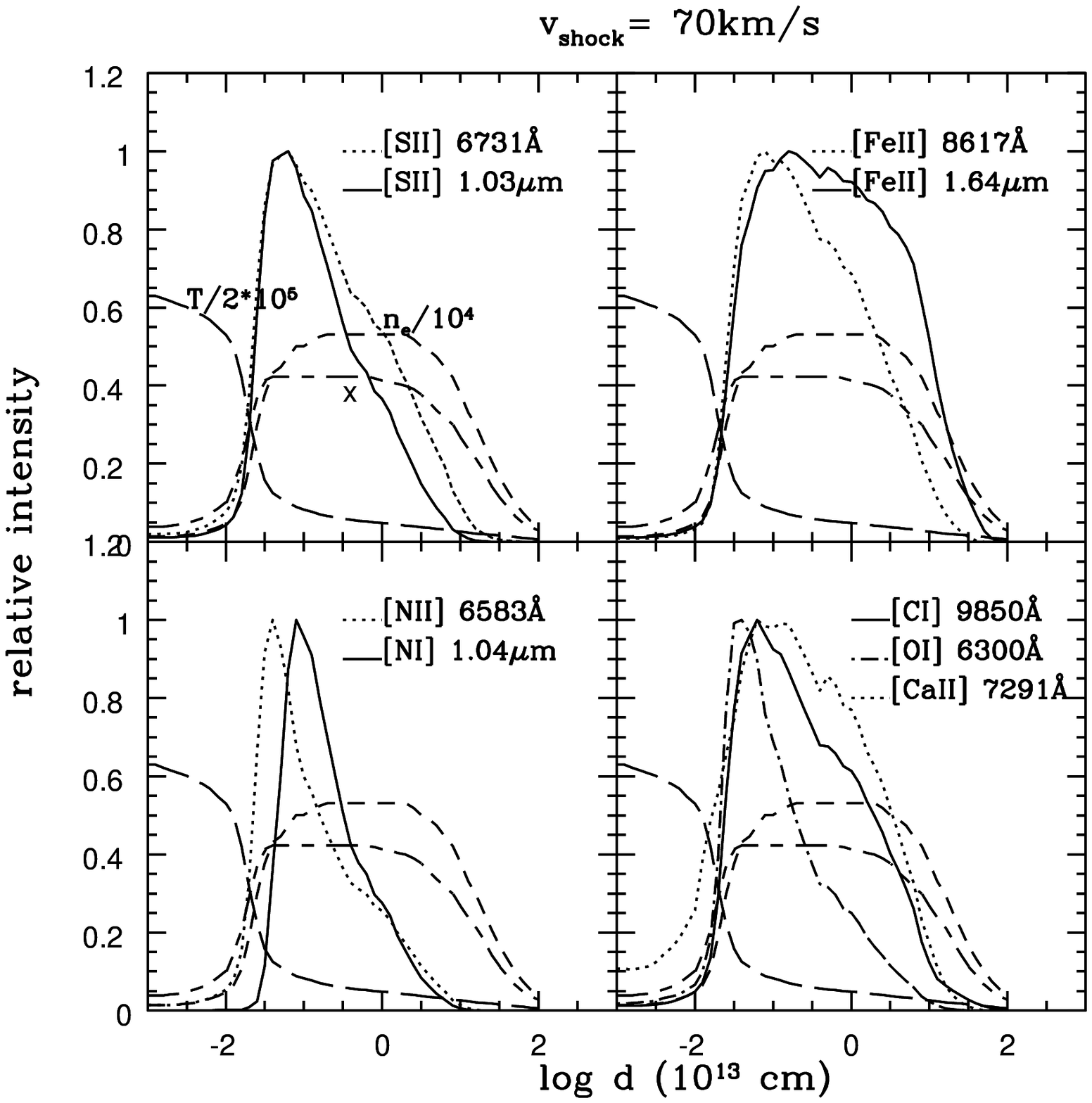}
   \caption{\label{fig:hart} Relative intensity profiles of several optical
   and infrared lines as a function of distance from the shock-front
   for a 70 km\,s$^{-1}$ shock. These profiles have been calculated
   taking the temperature, density  and ionization fraction profiles 
   from Fig. 1 of Hartigan et al. (1994).}
\end{figure*}

The observed spectral range includes transitions from neutral and singly
ionized atomic species which have different excitation temperatures and critical 
densities,
thus allowing us to probe regions with different conditions.
In HH jets the lines are believed to be excited in the cooling region 
behind a shock front. This zone, which is generally about 10$^{14}$ cm 
wide (i.e. 0$\farcs$013 at 460pc) is not spatially resolved in our 
spectra, but the various lines may originate from different parts of 
the post-shocked regions. This is shown
in Fig. \ref{fig:hart}, where we plot the emission profiles 
(normalized to their peak value) in the cooling region of a 70 \kms\, 
shock. The profiles of $x_{e}$, $T_{e}$, and $n_{e}$ used to compute 
the intensity profiles were taken from the shock model of Hartigan et al. (1994) 
(their Fig. 1). Similar plots
were shown by BE99, who noted that the optical
\s, [\ion{N}{II}], and [\ion{O}{i}] transitions originate from similar regions at intermediate
temperatures and ionization fractions. From our Fig. \ref{fig:hart} we see
that \s\, transitions at about 1.03$\mu$m come
from a more compact and hot region, as they have a higher excitation temperature 
with respect to the optical lines. 
On the other hand, the emission zone 
of the \fe 1.64$\mu$m line is broader and covers most of the post-shocked 
region, tracing mainly the gas at higher density and lower temperature. 
The profiles of other strong lines, such as [\ion{C}{i}]9850\AA\, and 
[\ion{Ca}{ii}]7291\AA\, 
are also indicated. In 
Appendix A, we describe in more detail the diagnostic capabilities 
of the various lines used for our analysis.

In addition to the atomic lines, we also detect H$_2$ emission, which is commonly 
observed to be spatially associated with the
atomic gas in jets from young stars. 
The H$_2$ lines in the HH1 jet all come from the first two vibrational
levels and thus trace temperatures of the order of 2000 K. The non-detection
of lines from higher vibrational transitions, together with the relatively
low reddening in at least the external sections of the jet,
allows us to infer that molecular gas at higher temperatures
contribute little to the overall flux (Giannini et al. 2004).
The origin of the molecular emission, closely associated with gas of
much higher excitation, is still not clear. The emission 
could originate in the post shocked layers where the gas has cooled down sufficently to 
become molecular again. Alternatively, it could be due to ambient material
put into motion by the passage of the atomic jet or to molecular gas
in the outer layers of a disk wind.

\section{Derived physical parameters along the HH1 jet}

Ratios of different lines can be used to constrain
physical parameters, and their variation, along the jet, as 
described in some detail in Appendix A. However, to apply a 
diagnostic procedure that combines lines that are relatively far apart in 
wavelength, as in this case, we need to determine accurately the 
extinction towards each knot, and to 
correct the various line fluxes accordingly. To this aim, we have used the \fe\, 
lines at 1.32 and 1.64\um, that arise from the same upper level. The 
reasons for our choice, instead of the more commonly used ratio
\fe1.64/1.25\um, and the procedure followed, are described in Appendix B. 
The derived $A_{V}$ values are listed in Table 2. 
There is a sharp decrease in the extinction, from 
$A_{V} \sim$8 to 1.5 mag from the inner to the outer knots.
Since for knots A to E the individual $A_{V}$ values fluctuate more 
than the associated errors, an average value was taken, with an 
uncertainty of 0.5 mag given by the dispersion of the different 
determinations.  A visual extintion around 1.5 mag in these outer knots is
 in agreement with the E(B-V) value of 0.4 found by Solf et al. (1988) towards
the HH1 bow shock, if a standard R=3 value is taken.

As described in Appendix A, the selected optical transitions from 
\s, [\ion{O}{i}] and [\ion{N}{ii}] were used to determine the 
electron density, electron
temperature, and ionization fraction in the knots adopting
the BE99 technique, which in 
turn allow us to derive the total density $n_{H}$=$n_{e}$/$x_{e}$ . 

Given the limited spectral resolution of the EFOSC2 
configuration, the \s\,doublet at 6716, 6731\AA\, was not resolved sufficiently in
all knots to allow us to
separately measure the line fluxes. In these cases, we used
data taken during a previous run with the same instrument but with higher
spectral resolution, that allowed us to separate the doublet 
(Medves, Bacciotti \& Eisl\"{o}ffel in prep.). We checked that 
the proper motion of the knots 
during the time interval between the two runs was negligible.

\begin{figure}[!ht]
\resizebox{\hsize}{!}{\includegraphics{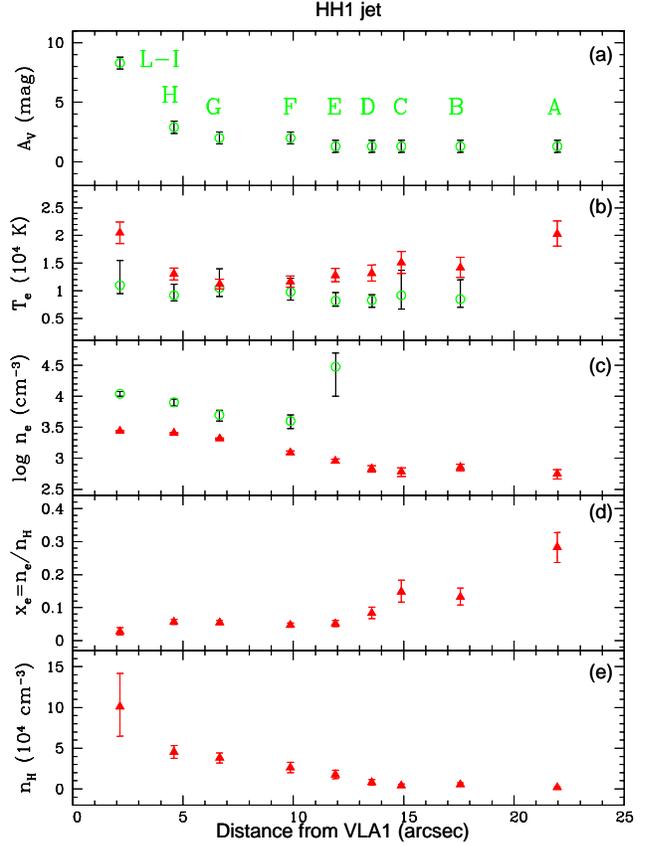}}
\caption{\label{fig:diag}
Derived physical parameters along the HH1 jet. Open circles 
refer to parameters estimated using diagnostics based on \fe\, lines, while
filled triangles refer to values determined through 
the BE99 technique. Error sources are described in Table 2.}
\end{figure}

The results of our analysis of the optical lines are shown in Tab. 3 
and Fig. \ref{fig:diag}, where the parameters are plotted
as a function of the distance from VLA1.
Moving outwards, we measure a decrease in the electron density 
(from $\sim$10$^3$ cm$^{-3}$ in knots L-I to $\sim$1.5\,10$^2$ cm$^{-3}$ in 
knots C-B), temperature (from 21000 K to 12000 K),
as well as in the total density (from 1.2\,10$^4$ cm$^{-3}$ 
to $\sim$2\,10$^3$ cm$^{-3}$). Such behaviour is often observed
in HH jets (BE99). 
The ionisation fraction $x_e$, on the contrary,  is  observed to slightly increase 
along  the jet. This may be due to the fact that material ejected 
earlier had a higher degree of ionization and lower 
electron density. In fact, for the conditions found in stellar jets the 
ionization fraction is not in equilibrium with the gas thermal 
conditions, since the time scale for hydrogen recombination, which is 
proportional to 1/$n_e$, is generally longer than the travel time along 
the bright section of the jet. Therefore, the ionization fraction is 
essentially  `frozen' into the gas, and may trace past events at large distances.
Alternatively, the gas may become re-ionized along the flow if, for 
example, the jet collides with an external clump (BE99)
 and an increase in ionization is induced by the UV photons
created in the shock (Molinari \& Noriega-Crespo 2002). This event, however,
seems unlikely in our case, because the resultant shock should produce 
an increase in the line fluxes which is not observed.

\begin{figure}[!ht]
\resizebox{\hsize}{!}{\includegraphics{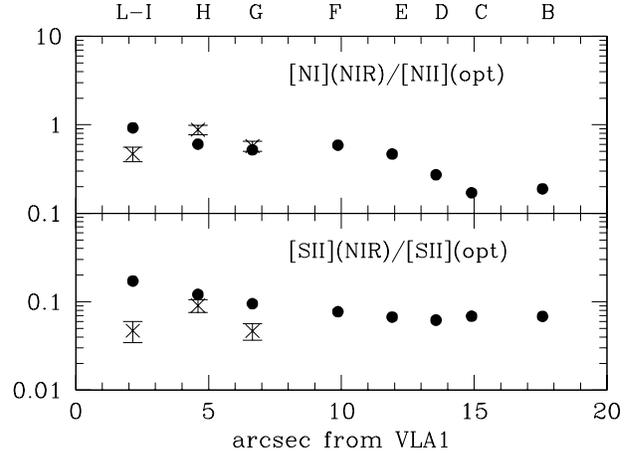}}
\caption{\label{fig:N-S} Comparison between predicted (filled circle)
and observed (crosses) dereddened ratios [NI]1.04$\mu$m/[NII]\lam\,6583 
(upper panel) and [SII]1.03$\mu$m/[SII]\lam\lam 6716,6731 (lower panel).
NIR lines were observed only in knots L-I, H and G. 
}
\end{figure}

The validity of our analysis can be checked by comparing the 
fluxes of other observed lines against the values predicted assuming the
derived physical parameters.
Fig. \ref{fig:N-S} shows such a comparison for the N and S
optical  and NIR lines.
The ratio \s(1.03$\mu$m)/\s(6716,6731\AA) indicates the presence 
of higher temperature components. The same 
information is given by [\ion{N}{I}](1.04$\mu$m)/[\ion{N}{II}](6583,6548\AA) 
which, in addition,
provides a check for the derived ionization fraction. 
In the BE99 diagnostics, in fact, the ionization fraction of hydrogen 
is found assuming that the O$^0$ and N$^+$ populations are regulated 
by recombination and charge-exchange with H. Thus the observed 
[\ion{N}{I}]/[\ion{N}{II}] ratio can be used to verify this
assumption. 
We see that for both [\ion{N}{I}]/[\ion{N}{II}]
and \s(IR)/\s(opt) the agreement between 
the predicted and observed values is quite good for knots H and
G while the predicted value is higher than the observed one 
for knots L-I. Since the same effect is seen in both line ratios, the problem does not arise 
from an overestimate of the 
hydrogen ionization fraction. 
Also, the presence of a temperature component
higher than that derived by our analysis would produce 
a discrepancy between the observed and predicted ratios even larger 
than the one we find. 
A better agreement between observed and predicted values in knots L-I 
can be obtained by assuming an $A_V$ value 
smaller than estimated, i.e. $\sim$6.5 mag instead of 8.3 mag. The agreement 
between observed and predicted 
ratio would also improve if the temperature in knots L-I is lower by 
about 60\% with respect to the estimated value. 


\begin{figure}
\resizebox{\hsize}{!}{\includegraphics{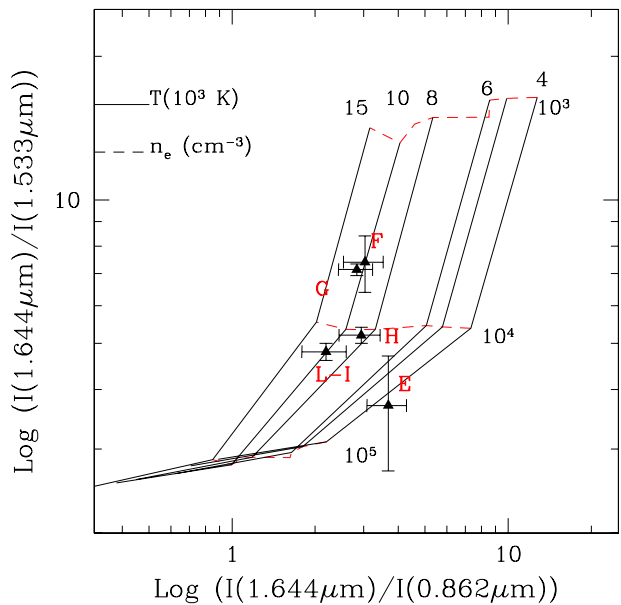}}
   \caption{\label{fig:fe_diag} Diagnostic diagram based on \fe\, line 
   ratios. The grid is constructed for electron densities of 10$^3$,
   10$^4$, and 10$^5$ cm$^{-3}$ (dashed lines), and for electron temperatures
   of 4,5,6,8,10,15$\times$10$^3$ K (solid lines). The symbols 
   indicate the line ratios observed in 
    the HH1 jet knots and dereddened assuming the $A_V$ values in Table 2. }
\end{figure}

\begin{figure}[!ht]
\resizebox{\hsize}{!}{\includegraphics{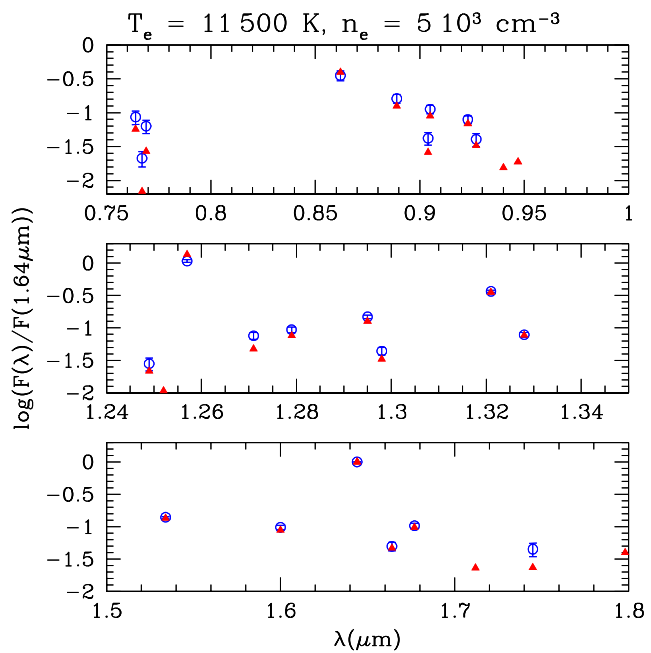}}
\caption{\label{fig:G_fit}
Dereddened \fe\, intensity ratios with respect to the 1.64\um\, line
 in knot G (open circles) against a model with 
$T_e$=11\,500 K and $n_e$=5\,10$^{3}$ cm$^{-3}$ (filled triangles). 
Intensities are normalized to the 1.64$\mu$m line.}
\end{figure}

The electron density and temperature can be
measured in a completely independent way from the ratios of specific \fe\, lines  
(see Fig. 
\ref{fig:fe_diag}). Our results
are also reported in Tab. 3 and plotted in Fig. \ref{fig:diag}.
In Fig. \ref{fig:G_fit}, we see that the \fe\, model
fits well all the observed \fe\, transitions in knot G. In fact
the majority of the line ratios are reproduced within about 30-40\%. 
Likewise, the $n_e$ determined from
optical and NIR \fe\, lines
decrease with distance from VLA1, however values derived using \fe\, are always 
a factor 3-4 higher with respect to those measured
from the \s\, lines. Such a result has already been found in the 
analysis of other HH objects (Nisini et al. 2002), and indicates that
the Fe emission originates from regions either denser and/or more ionized
than those giving rise to the optical lines. 
The electron temperatures derived from the \fe\, line
analysis are consistent with those derived from the optical diagnostics if
we consider the associated uncertainty. The average values of $T_e$(\fe), however,
are systematically lower than the $T_e$(opt) values.
In Fig. \ref{fig:hart} we see that such a 
behaviour is expected because of the different spatial distribution of Fe 
and optical lines in the cooling region of shocks. In fact, Fe lines
originate in a wide region of the post-shocked gas having, on average,
a lower temperature
and higher density than the zone producing the optical lines.

\begin{table*}
\caption[]{HH1 jet: physical parameters}

\vspace{0.5cm}
    \begin{tabular}[h]{ccc|cccc|cc}
      \hline \\[-5pt]
      &   & & \multicolumn{4}{c}{Diagnostics from O/S/N lines} & \multicolumn{2}{c}{Diagnostics from Fe lines} \\
Knot  & D$^{a}$ &  A$_{\rm V} \pm \Delta$A$_{\rm V}^{b}$ & n$_e \pm \Delta$n$_e$$^{c}$ & x$_e \pm \Delta$x$_e$$^{c}$
 & n$_{\rm H} \pm \Delta$n$_{\rm H}$$^{c}$ & T$_{e} \pm \Delta$T$_{e}$$^{c}$ & n$_e \pm \Delta$n$_e$$^{c}$ & T$_{e} \pm \Delta$T$_{e}$$^{c}$\\
      & $\arcsec$ & mag  & 10$^{3}$ cm$^{-3}$  &  & 10$^{3}$ cm$^{-3}$ & K &  10$^{3}$ cm$^{-3}$ & K\\[+5pt]
\hline \\[-5pt]
L-I & 1.2--3.8    &   8.3$\pm$0.5  & 2.8$\pm$0.1 & 0.02$\pm$0.01 &  101.1$\pm$40.6& 20\,500$\pm$1900 & 11$\pm$1 & 11000$^{+4500}_{-1500}$\\
H  &  3.8--5.5   &    2.9$\pm$0.5  & 2.6$\pm$0.1 & 0.06$\pm$0.01 &   45.4$\pm$7.9 & 13\,000$\pm$1100 &  8$\pm$1   & 9200$^{+2000}_{-1000}$ \\
G  &  5.5--8.5   &    2.0$\pm$0.5  &  2.1$\pm$0.1 & 0.05$\pm$0.01 &   38.1$\pm$6.1 & 11\,200$\pm$900 &  5$\pm$ 1   & 10\,500$^{+3500}_{-1500}$\\
F  &  8.5--11	 &    2.0$\pm$0.5  & 1.2$\pm$0.1 & 0.05$\pm$0.01 &   26.4$\pm$6.2 & 11\,700$\pm$1000 &  4$\pm$1    & 9800$^{+2500}_{-1500}$ \\
E  &  11--12.4   &    1.3$\pm$0.5  & 0.9$\pm$0.1 & 0.05$\pm$0.01 &   17.7$\pm$5.1 & 12\,800$\pm$1200 & 30 $\pm$ 20 & 8200$^{+1500}_{-1000}$ \\
D  &  12.4--13.8 &    1.3 $\pm$0.5  &  0.7$\pm$0.1 & 0.08$\pm$0.02 &	8.2$\pm$3.2 & 13\,200$\pm$1400 & ... & 8300$^{+1000}_{-1300}$ \\
C  &  13.8--15.6 &    1.3$\pm$0.5 &    0.6$\pm$0.1 & 0.15$\pm$0.03 &	4.1$\pm$1.9 & 15\,100$\pm$2000 & ... & 9200$^{+4500}_{-2500}$ \\
B  &  15.6--19.1 &    1.3$\pm$ 0.5  &  0.7$\pm$0.1 & 0.13$\pm$0.03 &	5.4$\pm$2.1 & 14\,200$\pm$1800 & ... &8500$^{+3500}_{-1500}$\\
A  &  19.1-26.2  &    1.3$\pm$0.5  &  0.6$\pm$0.1 & 0.28$\pm$0.05 &    2.0$\pm$0.8 & 20\,300$\pm$2300 & ... &...\\
\hline \\[+5pt]
      \end{tabular}

~$^{a}$Distance from VLA1 in arcsec\\
~$^{b}$Visual extinction measured from the ratio [FeII]1.64/1.32$\mu$m (see
Appendix 2 for details).\\
~$^{c}$n$_{\rm H}$ is calculated as x$_e$/n$_e$. Errors are estimated using 
both flux uncertainties and error propagation from uncertainties in the 
$A_{\rm V}$ determination and in solar abundance values reported in literature
(Asplund et al. 2004). 

\end{table*}

\subsection{Additional constraints on the excitation and physical structure}

\begin{figure}[!ht]
\resizebox{\hsize}{!}{\includegraphics{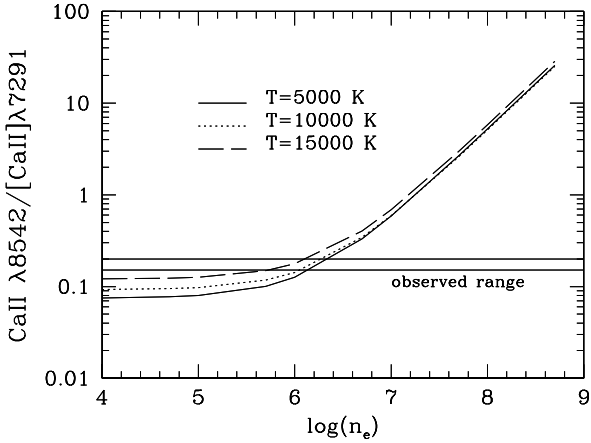}}
\caption{\label{fig:caII}
Calculated \ion{Ca}{ii}$\lambda$8542/[\ion{Ca}{ii}]$\lambda$7291 ratio, as a function
of the electron density and  temperature. The solid lines indicate the range of 
 values observed in the HH1 jet.
}
\end{figure}

As highlighted in Appendix A, several other line ratios 
can further constrain jet physical parameters. 
Our assumption of purely collisionally excited gas can be checked using line
ratios sensitive to fluorescence pumping effects. The dereddened 
[\ion{Ni}{ii}]$\lambda$7412/$\lambda$7378 ratio is one of them, with
expected values, for $n_e$ in the range 10$^2$ to 10$^5$ cm$^{-3}$,
 between 0.04 and $\sim$0.11 for pure collisions, and 
$\sim$0.11 and 0.4 if fluorescence excitation is included (Bautista et al. 1996).
We obtain for this ratio a value 
ranging from 0.04($\pm$0.02) in knot H to 0.14($\pm$0.06) in knot L-I,
indicating that the excitation is mostly collisional.
An additional probe of the
absence of fluorescence pumping is given by the 
[\ion{Ca}{ii}]$\lambda$7291/$\lambda$7324
ratio, which is expected to be 1.5 for collisional excitation
(Hartigan et al. 2004). Our dereddened ratios are always consistent 
with this value within the errors. 
This implies that the permitted \ion{Ca}{ii} transitions at 8542 and 8662\AA, 
detected in the inner knots E to L-I, are also excited by collisions. 
In turn, the ratio \ion{Ca}{ii}\lam8540/[\ion{Ca}{ii}]\lam7291 can be used as diagnostic 
of high electron density as shown in Fig. \ref{fig:caII}. This diagram
has been constructed adopting a five-levels statistical equilibrium code
that uses radiative transition rates from NIST, and collisional rates
from Mendoza (1983) and Chidichimo (1981).
This ratio is almost independent of the temperature, while it starts to
increase above $\sim$0.1 only for electron densities larger than
$\sim$10$^{6}$ cm$^{-3}$. The observed values thus 
imply $n_e$ $\sim$10$^{6}$ cm$^{-3}$.

\begin{figure}[!ht]
\resizebox{\hsize}{!}{\includegraphics{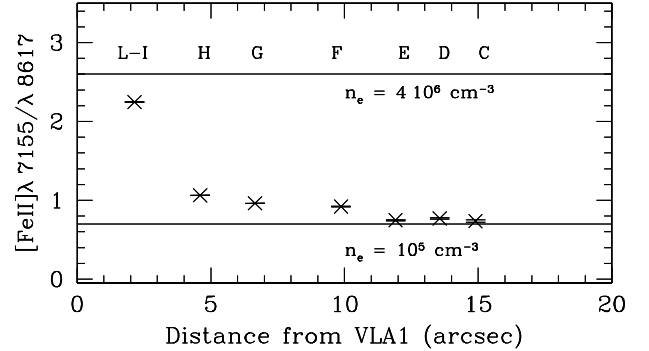}}
\caption{\label{fig:feII_hd}
Observed values of the ratio \fe\lam\,7155/\lam\,8617, which is a tracer
of high density regions. The two horizontal lines indicate the
value expected for densities of 10$^{5}$ and  6\,10$^{6}$ cm$^{-3}$ in a
collisionally excited gas at $T$=10000 K (Bautista \& Pradhan 1998).
}
\end{figure}

High electron densities along the jet are also indicated 
by the \fe$\lambda$7155/$\lambda$8617 ratio,  which, as described in 
Appendix A, is a tracer of regions at densities larger than $\sim$ 10$^{5}$
cm$^{-3}$.
Figure \ref{fig:feII_hd} shows the 
observed values against the predictions of Bautista \& Pradhan (1998),
for $n_{e}$ in the range 10$^{5}$--6\,10$^{6}$ cm$^{-3}$. Electron densities as 
high as 10$^{6}$ cm$^{-3}$ seem to be reached in knot L-I, but a 
component at $n_{e}$ $\sim$10$^{5}$ cm$^{-3}$ persists further out 
along the jet. Note that these values are about one order of magnitude higher than those
inferred from the NIR \fe\, lines.
 Values of electron density up to 10$^{6}$ cm$^{-3}$ have also been
inferred in the inner and denser regions of few T Tauri stars 
by Hartigan et al. (2004).

Finally, we point out the non-detection  of the
auroral optical [\ion{O}{ii}] transitions in any of the knots. 
The 7331\AA\, line pair
are never detected, while the 
doublet at 7321\AA\, could be  blended with [\ion{Ca}{ii}]7324\AA\,. 
Since, however,  the ratio 
[\ion{Ca}{ii}]\lam7291/\lam7324 is always close to the value expected from theory,
the contribution of [\ion{O}{ii}]\lam7321 should be negligible. This, in turn,
is independent confirmation that the hydrogen ionization fraction, which is strongly 
tied to the O$^+$/O$^0$ fraction by charge exchange, is low 
everywhere along the jet. In knot G, for example, the upper limit 
[\ion{O}{ii}]\lam7331/[\ion{O}{i}]\lam6300
$\sim$8\,10$^{-3}$ derived at the 3$\sigma$ level, 
implies an upper limit for the
hydrogen ionization fraction $x_{e}$$\sim$0.1 (Hartigan et al. 2004), 
which is consistent with $x_e$=0.05 derived by us. 

\section{Mass flux in the jet}

The knowledge of the physical conditions along the jet beam allows us
to derive the mass flux carried by the HH1 jet, which is a fundamental 
quantity regulating the dynamics of the accretion/ejection
process in protostars.

For the determination of \.{M} we exploit two different procedures.
In the first method, we directly estimate the mass flux in each knot
combining the total density along the jet beam with values of
jet radius, $r_J$,  and velocities, $v_J$ available in the literature
($\dot{M} = \mu\,m_H\,n_H\times\pi\,r_J^2\,v_J$,
where $m_H$ is the proton mass, and $\mu$=1.24 is the average atomic weight). 
This method is not affected by uncertainties in the 
distance of the object and reddening estimate.
On the one hand it assumes that the knot is uniformely filled at 
the density derived from the diagnostics, giving an upper limit to 
$\dot{M}$. On the other hand, such an effect is 
partially compensated for by the presence of regions at even higher densities
in the beams than those traced by the \s\, lines. 

\begin{figure*}[!ht]
\sidecaption
\includegraphics[width=12cm]{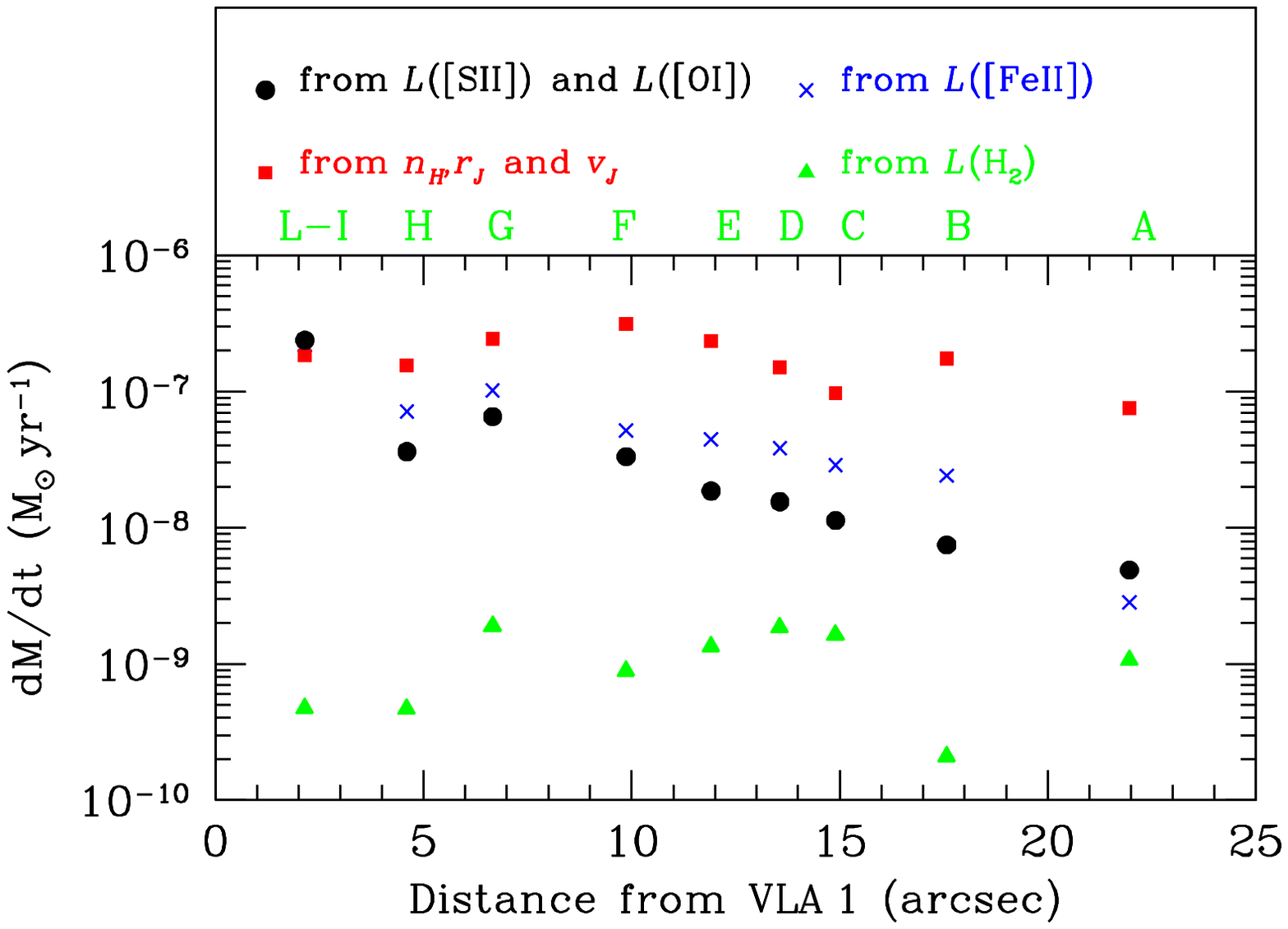}
\caption{\label{fig:mdot} Comparison of the \.{M} determinations 
along the HH1 jet derived adopting different methods and lines. Filled
squares refer to \.{M} values directly measured from our derived total
density and the knot radius taken from Reipurth et al. (2000),
assuming complete beam filling. Filled circles are mass loss rates
estimated from \s$\lambda\lambda$6716,6731 and [OI]$\lambda$6300 luminosities.
The two values coincide inside the symbol. Finally, crosses and triangles refer
to the \.{M} value derived from the \fe 1.64$\mu$m and H$_2$ 2.12$\mu$m 
luminosities.
}
\end{figure*}

The mass flux derived with this method is given in
Table 3 and plotted in Fig \ref{fig:mdot}. For r$_{J}$, we have
taken half of the FWHM of the \s\, intensity profile in HST images
measured by Reipurth et al. (2000), which increases from 0.1 to 1 arcsec
along the jet. Jet velocities are derived from proper motion studies
by Bally et al. (2002), assuming an inclination angle of 10$^{o}$ (see Table 3).
The derived mass flux ranges between 1 and 3\,10$^{-7}$ M$_{\odot}$\,yr$^{-1}$
and is fairly constant along the jet within the uncertanties.

The second method to derive the mass flux uses the observed luminosities of 
selected lines such as \s, [\ion{O}{I}] and \fe. These lines are optically 
thin, thus their luminosity is proportional to the mass of the emitting gas 
(see e.g. Hartigan et al. 1984). We have:
\begin{equation}
\dot{M} = \mu\,m_H\times(n_H\,V)\times v_{t}/l_{t}
\end{equation}
\begin{equation}
n_H\,V = L(line)\,\left(h\,\nu\,A_{i}\,f_{i}\,\frac{X^i}{X}\,\frac{[X]}{[H]}\right)^{-1}
\end{equation}
where  $A_{i}$, $f_{i}$ are the radiative rate and fractional population 
of the upper level of the considered transition, 
$X^i/X$ is the ionization fraction of the species $X$
having total abundance with respect to hydrogen $[X]/[H]$, 
$v_{t}$ and $l_{t}$ are the  velocity and length of the knot, 
projected perpendicular to the line of sight.

This method is affected by uncertainties in absolute 
calibrations, extinction, and distance, but does 
implicitly take into account the volume filling factor. In practice, the latter is
given by the ratio of $\dot{M}$ values derived from the two 
different methods.
We have computed $\dot{M}$ from the luminosity of
\s$\lambda\lambda$ 6716,6731, [\ion{O}{I}]$\lambda$6300, and \fe 
1.64$\mu$m lines, assuming all Fe is in gaseous form (Tab. 4, Fig.\ref{fig:mdot}).
Since, as it will be discussed in section 4.4, up to $\sim$70\% of Fe atoms 
can still be locked into grains, the derived $\dot{M}$(\fe)
is actually a lower limit to the real value. 

\begin{table*}
\caption[]{Mass flux in the jet and filling factors}

\vspace{0.5cm}
    \begin{tabular}[h]{cccccccc}

      \hline \\[-5pt]
knot & $v_{\rm tan}$$^{a}$ & $r_{jet}$$^{b}$ & $\dot{M}$($n_{\rm H},r_{j},v_{j}$)$^{c}$ 
&$\dot{M}$(\s)$^{d}$ & $ff$(\s)$^{e}$ & $\dot{M}$(\fe)$^{d}$ &  $\dot{M}$(H$_2$)$^{f}$\\

     & km\,s$^{-1}$  & arcsec  & M$_{\odot}$\,yr$^{-1}$ &M$_{\odot}$\,yr$^{-1}$ & &M$_{\odot}$\,yr$^{-1}$
    & M$_{\odot}$\,yr$^{-1}$  \\[+5pt]
\hline \\[-5pt]
L-I & 300 &  0.1  &  1.8\,10$^{-7}$ & 2.4\,10$^{-7}$ & 1.3 & 2.2\,10$^{-7}$  & 4.7\,10$^{-9}$\\
H  &  300 &  0.15 &  1.6\,10$^{-7}$ & 3.6\,10$^{-8}$ & 0.2 &  7.1\,10$^{-8}$ &   4.7\,10$^{-10}$\\
G  &  305 &  0.2  &  2.4\,10$^{-7}$ & 6.9\,10$^{-8}$ & 0.3 &  1.0\,10$^{-7}$ &   1.9\,10$^{-9}$\\
F  &  300 &  0.3  &  3.1\,10$^{-7}$ & 3.7\,10$^{-8}$ & 0.1 &  5.2\,10$^{-8}$ &   8.9\,10$^{-10}$\\
E  &  281 &  0.3  &  2.3\,10$^{-7}$ & 1.8\,10$^{-8}$ & 0.08&  4.4\,10$^{-8}$ &   1.3\,10$^{-9}$\\
D  &  300 &  0.35 &  1.5\,10$^{-7}$ & 1.5\,10$^{-8}$ & 0.1 &  3.8\,10$^{-8}$ &   1.8\,10$^{-9}$\\
C  &  296 &  0.4  &  9.7\,10$^{-8}$ & 1.1\,10$^{-8}$ & 0.1&  2.9\,10$^{-8}$  &   1.6\,10$^{-9}$\\
B  &  317 &  0.5  &  1.7\,10$^{-7}$ & 7.5\,10$^{-9}$ & 0.04&  2.4\,10$^{-8}$ &   2.1\,10$^{-10}$\\
A  &  311 &  0.5  &  7.6\,10$^{-8}$ & 5.1\,10$^{-9}$ & 0.07&  2.8\,10$^{-9}$ &   1.1\,10$^{-9}$\\
\hline \\[+5pt]
      \end{tabular}

~$^{a}$ taken from the proper motion study by Bally et al. (2002). Total velocities
($v_{jet}$) have been derived from these values assuming an inclination angle of 10$^{0}$.\\
~$^{b}$ taken from the FWHM intensity transverse profiles in HST images of Reipurth et al. (2000)\\
~$^{c}$ $\dot{M}$($n_{\rm H},r_{j},v_{j}$) = $\mu\,m_H\,n_H\times\pi\,r_{jet}^2\,v_{jet}$, where $n_H$ is the total
density reported in Table 2. Volume filling factors equal to unity are assumed.\\
~$^{d}$ $\dot{M}$ measured from the luminosities of the \s\lam\lam\,6716,6731 lines and \fe 1.64$\mu$m line.\\
~$^{e}$ Volume filling factor estimated from the ratio between the observed and predicted line luminosities
(see text). They are also equal to the ratio between $\dot{M}$ and
$\dot{M}$($n_{\rm H},r_{j},v_{j}$) derived from line luminosities.\\
~$^{f}$ $\dot{M}$ measured from the luminosity of the H$_2$ 2.12$\mu$m line, assuming the H$_2$ temperature
estimated from the ratio of the different H$_2$ observed lines.

\end{table*}

Fig. \ref{fig:mdot} illustrates the \.{M} values obtained with
the different methods. We note that the values derived from line luminosities
are lower than those calculated from the jet density and radius, and  
fall off along the jet, indicating a decrease of filling factor.
In fact, measuring the volume filling factor ($ff$) by
comparing the observed luminosity with that expected from the considered 
volume, one finds that $ff$ varies from $\sim$1 to $\sim$0.06 from knot L-I to knots A-B
(Table 3).
This behaviour is not unexpected, as the furthest knots have a larger 
volume 
filled with more diffuse gas, while  the knots closer to the star 
are more dense and compact. 

In Fig. \ref{fig:mdot} we note that the mass flux derived from \fe\, line luminosity is
larger than $\dot{M}$(\s,[\ion{O}{i}]), at least from knot H outwards. This
suggests that the higher $n_{e}$ values, given by 
the \fe\, diagnostic,  come from gas with higher average density 
than in the \s\, line emitting region, rather than from gas with a higher 
ionization fraction. Thus the \fe\, 
luminosity is probably a better tracer of the mass flux than the
[\ion{O}{i}] or \s\, line luminosity, since the former traces a
larger portion of the total mass flowing through a single knot. 
\\

\.{M} derived from line luminosities, which takes into account the filling factor,
 decreases with distance from the source. 
A constant mass flux along the jet
is expected if the flow is stationary, since the decrease in 
the total density should be compensated for by the larger jet radius. 
A mass flux decrease can be due to a change in the jet physical conditions: 
if the gas recombines and the temperature decreases, the bulk of the flowing mass may be 
mainly in neutral atomic or molecular form and thus not be traceable through
its ionic emission. 
To gauge the mass flux associated with the molecular components we have determined
$\dot{M}$(\h) using the 2.12\um\, line luminosity. The temperature of the
warm \h\, gas is 2000--2500 K, as derived from the ratios 
of the dereddened line fluxes. Then, the total N$_{H2}$ column density is measured
by comparing the intrinsic (per unit mass) and observed flux of the 1-0 S(1) 2.12$\mu$m line. 
 The \h\, tangential velocity has been assumed equal to that derived from optical lines,
following Noriega-Crespo et al. (1997), who measured \h\, knot proper motions
comparable to those obtained from atomic emission lines. 

Our values of $\dot{M}$(\h\,) range between 2\,10$^{-10}$ and 2\,10$^{-9}$
M$_{\odot}\,yr^{-1}$, and are therefore about 2-3 orders of magnitude
lower than those derived from the atomic lines. Moreover,
\.{M}(H$_2$) does not increase with distance. Thus it is unlikely that the 
decrease of the mass flux derived for the atomic/ionic component 
is due to atomic gas turned into molecular material.
Other contributions, however,
such as atomic gas at low temperature, not emitting in the optical and near IR, 
may become important to the bulk of mass flow.
In this regard, we note that strong [OI]63$\mu$m line emission, which
traces shocked atomic gas at temperatures below 5000 K,  has been
detected by ISO in a 70$\arcsec$ beam centred on VLA1 (Giannini et al. 2001,
 Molinari \& Noriega-Crespo 2001).
Similar 63$\mu$m emission has also been found towards HH1 and HH2, 
suggesting that such emission is not excited in the source 
circumstellar envelope, but derives from both the
shocked gas and the diffuse PDR of the hosting cloud. This latter 
contribution has been estimated to account for about 50\% of the total [OI] luminosity.
Excluding the PDR contribution and estimating the mass flux from the [OI]63\um\,
luminosity (Hollenbach 1985), we derive $\dot{M}$([OI]63\um) $\sim$
3\,10$^{-6}$ M$_{\odot}\,yr^{-1}$. Such a large value suggests that a
significant part of the mass flux may be due to a warm atomic component.
FIR observations at resolution higher than that available with ISO, may be obtained in 
the near future with SOFIA and Herschel, and will allow us to gain a more complete
picture.
\\

Finally, we note that $\dot{M}$(\h) in the HH1 jet is on average
lower by one or two orders of magnitude with respect to the values 
derived by Davis et al. (2001, 2003) for other small-scale \h\, jets
from embedded sources. These authors also find that 
$\dot{M}$(\h) and \.{M}(FeII) are roughly comparable for the majority of the
objects in their sample. In contrast, our estimates for HH1 show its 
this jet, despite its origin from an embedded source, presents characteristics
similar to jets from more evolved YSOs. In such sources 
the molecular component contributes little to the jet 
excitation and dynamics in comparison to the atomic component.

We also note that the molecular outflow associated with the VLA1 source
is weak and less energetic than other outflows from Class 0 sources
(Moro-Mart\'{\i}n et al. 1999). The momentum flux deduced by our observations
of the atomic jet ($\sim$3\,10$^{-5}$ M$_{\odot}\,km\,s^{-1}\,yr^{-1}$)
is comparable to that estimated for the associated CO outflow (Correia
et al. 1997). All thi evidence suggests that around VLA1 
the density of molecular material is low and thus that the dynamics
of the flow is mainly determined by the atomic gas.

\section{Abundances of refractory elements}

The observed wavelength range includes several lines from refractory
species. In addition to Fe, we detected transitions from C, Ca,
Ni and Cr. These species are expected to be largely depleted in neutral or
molecular interstellar regions, while their gas phase abundance is enhanced
in shocked regions where the dust grains have been at least partially destroyed 
either by 
sputtering or by photoevaporation (e.g. Draine 2003). 
Therefore, the determination of gas phase abundances
can provide constraints on dust properties and structure, on dust destruction 
mechanisms and on how such mechanisms depend on physical parameters such 
as jet velocity
and excitation conditions. In this regard we note that the different species are expected 
to follow selective patterns for their erosion from the dust grains, and thus
the fraction of their gas-phase abundance can be different and give further 
indications of dust grain reprocessing.

To derive quantitatively the gas-phase elemental abundance, however, one has to rely on 
comparing emission of refractory and non-refractory species
that originate in the same region. In this way, the difference between predicted ratios of lines,
calculated with the determined values of $T_e$, $n_e$ and $x_e$, and the observed ratios
gives indications on the depletion of the refractory species with respect to the
assumed solar abundance. For Fe this procedure is quite 
complicated, and contrasting results have been reported for HH objects 
(see e.g. B\"ohm \& Matt 2001, Beck-Winchatz et al. 1996, Nisini et al. 2002).
The fact that the Fe spectrum is consistent with physical conditions different from those
inferred from the optical lines (e.g. \s\lam6716 and [\ion{O}{i}]\lam6300) gives
strong limitations to a diagnostics based on a comparison between such lines.

\begin{figure}[!ht]
\resizebox{\hsize}{!}{\includegraphics{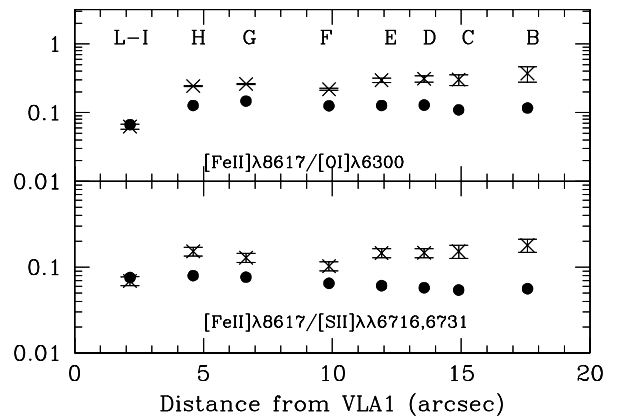}}
\caption{\label{fig:Fe_O_S}
Ratios of \fe8617\AA\, line with respect to 
\s\lam\lam\,6716,6731 
and [\ion{O}{i}]\lam6300 in the various knots. Crosses: measured 
de-reddened values. Filled circles: predicted values assuming $T_e$ and $n_e$ 
derived from the BE99 analysis. Fe and S have been assumed completely ionized once,
while the O$^+$/O$^0$ fraction has been calculated using the ionization model
for this species. }
\end{figure}

This is demonstrated in Figure \ref{fig:Fe_O_S}, which is a plot of the
line ratios  \fe$\lambda$8617/\s$\lambda\lambda$6716,6731 and 
\fe$\lambda$8617/[\ion{O}{i}]$\lambda$6300, assuming a Fe gas-phase abundance equal to  
solar (3\,10$^{-5}$, Grevesse \& Sauval 1998).
The \fe$\lambda$8617 line has been chosen because its spatial distribution is similar 
to the distribution of the \s\, and [\ion{O}{i}] lines, as shown in Figure 3. 
However, Figure \ref{fig:Fe_O_S} shows that the predicted ratios, assuming 
for all lines the physical conditions derived from the BE99 analysis, 
underestimate the observed values and, in fact, would suggest Fe gas phase 
abundances {\it larger} than solar. 

\begin{figure}[!ht]
\resizebox{\hsize}{!}{\includegraphics{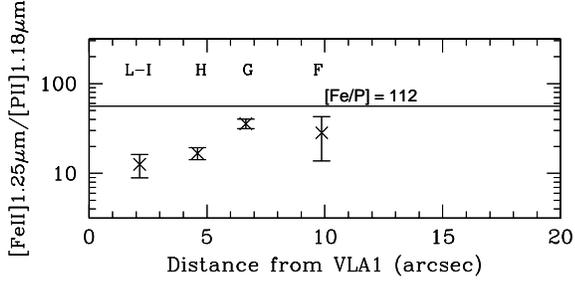}}
\caption{\label{fig:Fe-P} Observed \fe1.25$\mu$m/[PII]1.18$\mu$m ratios in
knots L-I, H, G and F. The line represents the predicted value expected
if all the iron is in gaseous form and assuming a [Fe]/[P] solar abundance
ratio of 112.
}
\end{figure}

We have then considered the ratio \fe1.25$\mu$m/[\ion{P}{ii}]1.18$\mu$m
(Figure \ref{fig:Fe-P}). 
Phosphorus is a non-refractory species
and its 1.18$\mu$m line has about the same excitation conditions as the 
\fe1.25$\mu$m line. As shown in Oliva et al. (2001), 
the ratio \fe1.25/[\ion{P}{ii}]1.18 should be about [Fe/P]/2. 
Assuming solar abundances ([Fe/P]$\sim$ 112) we find [Fe]$_{gas}$/[Fe]$_{solar}$ 
$\sim$0.7 for knots G and F and $\sim$0.2-0.3 for the internal knots L-I and H, in line
with values derived in Nisini et al. (2002) for other HH objects.
Finally, a Fe gas-phase abundance lower than solar is also inferred from a comparison of 
\fe\, lines and Pa$\beta$ as in Nisini et al. (2002). In knot G, taking a 
Pa$\beta$ upper limit 
of 5\,10$^{-16}$ erg\,s$^{-1}$\,cm$^{-2}$, $T_e$=11\,000 K and $x_e$=0.05,
we derive a lower limit of [Fe]$_{gas}$=0.1\,[Fe]$_{solar}$ from the plot in Fig. 13 
of Nisini et al. (2002). 
\\

\begin{figure}[!ht]
\resizebox{\hsize}{!}{\includegraphics{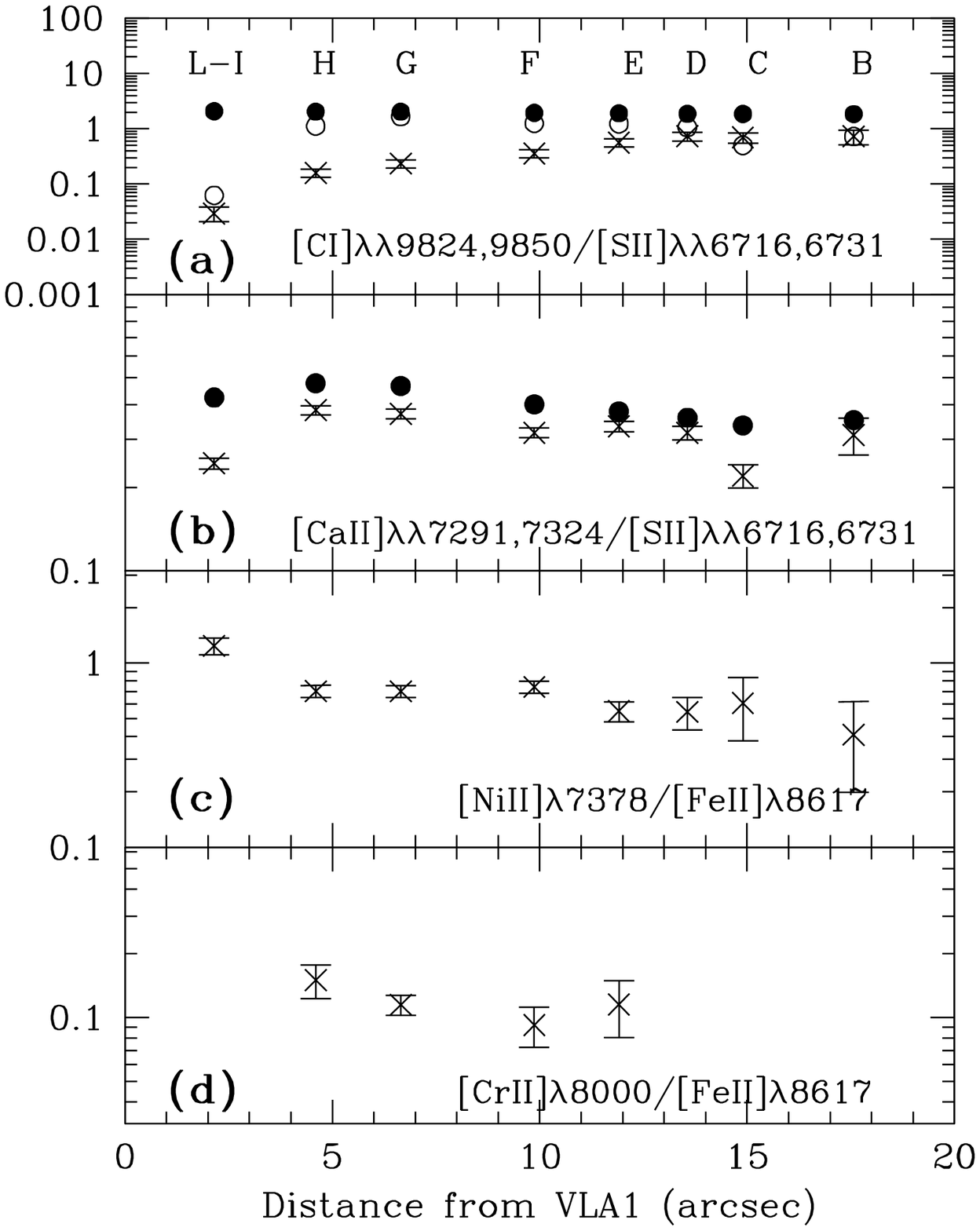}}
\caption{\label{fig:Ca-C}
Dereddened ratios of lines from refractory species (crosses): (a)
 [\ion{C}{i}]\lam\lam\,9824,9850/\s\lam\lam\,6716,6730,
(b)  [\ion{Ca}{ii}]\lam\lam\,7291,7324/\s\lam\lam\,6716,6730,
(c)   [\ion{Ni}{ii}]\lam7378/\fe\lam8617,
(d) [\ion{Cr}{ii}]\lam8000/\fe\lam8617.
In panels (a) and (b), filled circles are the predicted values assuming 
Ca, C and S have solar abundances, C all neutral and Ca, S all singly ionized.
Open circles in panel (a) refer to predictions where the 
 C$^0$/C$^+$ abundance ratio has been computed through ionization equilibrium analysis.}
\end{figure}

The gas phase abundance of C and Ca is more easily derived from a comparison 
with the \s\, optical lines assuming the same excitation conditions (see Fig. 3).
Figure \ref{fig:Ca-C} shows the comparison between the observed and predicted 
ratios [\ion{C}{i}]$\lambda\lambda$9824,9850/[\ion{S}{ii}]$\lambda\lambda$6716,6731 and 
[\ion{Ca}{ii}]$\lambda\lambda$7291,7324/\s$\lambda\lambda$6716,6731.

While Ca can be assumed to be completely ionized, the fraction C$^{0}$/C$_{tot}$ 
has been computed with an ionization balance between collisional 
and charge-exchange ionization, and direct/dielectronic recombination
(rates from Stancil et al. (1998) and Landini \& Monsignori Fossi (1990)).
Figure \ref{fig:Ca-C} shows that the observed [\ion{C}{i}]/\s\, ratio
decreases towards the source, in contrast to what can be expected theoretically.
In knots L-I, however, the higher temperature causes C to become
almost completely collisionally ionized and the theoretical ratio is also
reduced.
The discrepancy between observed and expected [\ion{C}{i}]/\s\, ratios
 can be interpreted as due to a 
decrease in the C gas-phase abundance. This behaviour can be 
correlated to the  
increase of the total density from about 5\,10$^{3}$ to 1.2\,10$^{5}$
cm$^{-3}$ going from knots A-C to knots L-I, since the efficiency
of depletion increases with the density. 
On the other hand, the higher temperature of knot L-I may cause dust 
evaporation of C to compete against depletion. 

In Fig. \ref{fig:Ca-C} we also show that the observed [\ion{Ca}{ii}]/\s\,
ratio is smaller than predicted for knots
F to L-I, although Ca depletion is lower than for C.
To check if selective depletion is present among other species, we have
plotted in Figure \ref{fig:Ca-C} the [\ion{Ni}{ii}]\lam7378/\fe\lam8617 and 
[\ion{Cr}{ii}]\lam8000/\fe\lam8617 ratios observed in the various knots. 
There is evidence that the [\ion{Ni}{ii}]\lam7378/\fe\lam8617 ratio decreases 
with distance. Since Ni$^+$ and Fe$^+$ coexist, and the two lines
have the same excitation conditions (Bautista et al. 1996), we interpret the
observed decrease as a change in the [Ni]/[Fe] gas abundance ratio.  
 In this respect it is interesting to note that Beck-Winchatz et al. (1996)
found a Fe abundance close to solar in the HH1 bow shock, but here the 
Ni/S ratio was 10 times the solar value.

The presented analysis, therefore, shows that a certain degree of depletion
of the refractory elements is  present, an effect which is apparently 
selective, being higher for Fe and C. This shows 
that dust grains are not completely destroyed in stellar jets, and that 
the number of refractory atoms released in the gas phase is an inverse
function of the total density. 
In fact, models for dust reprocessing in shocks show that shattering processes, 
and grain-grain collisions, are able to destroy 
only a few percent of large grains for shock velocities $v_s \ga$ 10 \kms\, 
(Draine 2003). At the same time, sputtering in fast shocks produces 
 complete dust destruction for $v_s >$ 100\kms\, (Jones 2000). 
Since shocks in stellar jets, like the HH1 jet, have velocities no larger 
than 40-50\kms,
we would expect that a large fraction of dust grains survive shocks
and contribute to the observed partial depletion of
refractory elements. Moreover, our results also show that 
multiple low velocity shock events, such as those occurring
in protostellar jets, are not 100\% efficient in completing the
dust destruction process.


\section{Summary}

We have presented a new combined optical/NIR analysis of flux calibrated spectra 
 of the HH1 jet. The simultaneous use of several diagnostic lines has 
 proven to be 
 very powerful in probing the density and temperature stratification
inside the jet knots, and has allowed us to obtain
a very detailed picture of the jet physical structure and excitation
even from moderate spectral and spatial resolution observations.

Using different line ratios, we have derived visual extinction, electron density, 
temperature and ionization 
fraction as a function of distance from the driving source.
The electron density from the \s\, doublet decreases from 3\,10$^{3}$ 
cm$^{-3}$ in the L-I knots (at d$<$4$''$ from VLA1) to $\sim$600 cm$^{-3}$
in the A-D knots far from the star. Conversely, $n_e$ values between 
10$^{4}$  and 4\,10$^{3}$ cm$^{-3}$
are derived in the same knots from \fe\, line IR ratios. Finally,
the use of very high density probes, such as \fe\,\lam7155/8617 and 
\ion{Ca}{ii}\lam8542/[\ion{Ca}{ii}]\lam7291,
supports the presence of jet components at even higher electron densities, 
ranging from 10$^{5}$ to 10$^{6}$ cm$^{-3}$. 
Such a stratification presumably originates in the different layers 
of a post-shocked gas. This is reasonable since 
the post-shock cooling distance is expected to range between $\sim$10$^{14}$ 
and 10$^{16}$ cm (Hartigan et al. 1987, see also Fig. 1), i.e. smaller
than our spatial resolution.
Our conclusion is also supported by the temperature
values inferred from the \fe\, optical/NIR ratios, which are lower in 
comparison to those
derived from using optical line diagnostics. Fe emission therefore 
traces post-shocked regions located at larger distances from the shock front
 than \s\, lines, where the compression is higher and the 
 temperature is declining.  

The ionization fraction, derived using the BE99 diagnostics, is rather 
low ($x_e \sim$0.02-0.1) which is also supported by the non-detection of the [OII] auroral 
lines at 7321 and 7331\AA. 
Several line ratios have been also used to exclude the presence of fluorescence
excitation even in the knots closest to the source. This implicitly validates
the diagnostic procedures we use, which assumes that the gas radiative properties
are regulated by collisional excitation.

If we assume that the ionization fraction is almost constant throughout most 
of the post-shocked line emitting region, as the profile in Fig. 1 would suggest, 
we deduce, in the innermost jet region, a total density as 
high as 5\,10$^{7}$ cm$^{-3}$. 
Moreover, the range of total densities probed by our analysis implies that in each knot 
a compression factor {\it C} = $n_{\rm postshock}/n_{\rm preshock}$ of at least 10$^2$ is reached.
For comparison, values of {\it C} lower than $\sim$100 are derived from 
dissociative shocks with velocity ranging from 30 to 100 \kms\, (Hartigan et al. 1994). 

The mass flux in the jet has been determined for each knot using two methods. In the 
first $\dot{M}$ is calculated combining the estimated 
$n_H$ values with the jet radius and velocity taken from the 
literature ($\dot{M}$({$n_H,r_{j},v_{j}$})).
In the second we used a comparison of observed and predicted total luminosity of the
\s, [\ion{O}{i}], and \fe\, lines. 
In the jet section closest to the source (i.e. knots L-I) 
both methods suggest 
a value of $\sim$2\,10$^{-7}$ M$_{\odot}$\,yr$^{-1}$.
Conversely, as we move away from the driving source we find discrepancies 
between the results from the two methods. The $\dot{M}$({$n_H,r_{j},v_{j}$}) 
value represents an upper limit
to the mass flux, since one makes the assumption that the jet  
is completely filled by 
gas at the estimated density. Conversely, $\dot{M}$([SII],[OI]) calculated from line
luminosities, takes into account the proper filling factor, but 
underestimates the real $\dot{M}$ value, since in this case one does not include the higher
density gas regions. $\dot{M}$([FeII]) is intermediate between these two values, and
 is probably a better tracer of the global mass flux through the jet.

$\dot{M}$([FeII] is found to decrease with the distance from the driving source,
from 2\,10$^{-7}$ in knots L-I down to $\sim$3-5\,10$^{-9}$ M$_{\odot}$\,yr$^{-1}$ in knot A. 
Such a decrease
may indicate that in the more distant regions of the jet an atomic component at $T<5000$ K 
(as traced by the [OI]63$\mu$m emission observed by ISO) becomes important and
contributes to the bulk of the mass flow.
The mass flux due to the warm molecular H$_2$ gas has also been estimated
and it turns out to be always smaller than the other values by one or two orders
of magnitude.

Finally, the presented observations suggest that the refractory elements
are still partially depleted with respect to the assumed solar abundance
by fractions which varies from less than 10\% in the low density regions up to
 $\sim$90\% for C, $\sim$70\% for Fe and 
$\sim$50\% for Ca in the inner and densest knots. This is a result expected in relatively low-velocity
shocks, since the complete destruction of dust by shattering processes is 
predicted only for shock velocities larger than 50-100 \kms. It also indicates
that the action of several low-velocity shock events during the jet lifetime 
is not effective in producing a complete destruction of dust grains. 
While we found that there is evidence of depletion depending on 
gas density, an analysis performed on a larger number of jets is needed 
to properly determine the dependence of the gas-phase abundances of refractory elements 
on jet excitation conditions. In addition, the observation of a selective pattern
for the depletion of different elements needs to be investigated further 
since it may provide important information on dust grain composition and properties.

\appendix
\section{Diagnostic capabilities of the observed lines }

\subsection{Oxygen, Nitrogen and Sulphur lines}

The forbidden lines from O, N, S in the optical range are powerful 
diagnostic tools for stellar jets, as discussed in BE99, Dougados et al.
(2000), Hartigan et al. (2004). In particular, the ratio of 
\s\lam6716 and \lam6731 is sensitive to electron densities in the range 
50-2\,10$^{4}$ cm$^{-3}$, this latter value being the critical density 
for collisional excitation. Furthermore, BE99 have developed a method to 
derive the hydrogen ionization fraction $x_{e}$ and the electron 
temperature T$_{e}$ from the observed ratios 
[\ion{O}{i}]\lam\lam6300,6363/[\ion{N}{ii}]\lam\lam6583,6548 
and [\ion{O}{i}]\lam\lam6300,6363/\s\lam\lam6716,6731. 
The technique is based on the fact 
that in typical conditions for the beams of stellar jets the hydrogen ionization 
fraction is primarily determined by charge-exchange with oxygen and 
nitrogen atoms, while collisional ionization of H and photoionization 
do not make  a large contribution until the shock velocity exceeds 100 
\kms. Such a velocity, however, is only reached in the large 
bow-shocks at the head of the jets, while those
in the beam are generally not larger than 30-40 \kms. 
An additional assumption is that S is all singly ionized. Such an 
assumption is motivated by the large I.P. of S$^{+}$ (50 eV) and by the
non-detection of the [\ion{S}{iii}] and [\ion{S}{i}] lines at 9533\AA\, and 1.082$\mu$m
respectively, in any of our spectra.
More details 
can be found in BE99 and Bacciotti (2002). The jets analysed 
with this technique in previous studies typically have
0.01$< x_{e} <$0.6 and 7000$< T_{e} <$2\,10$^{4}$. When using the 
BE technique a set of elemental abundances has to be assumed. In 
this study we have adopted solar abundances from Grevesse \& Sauval 
(1988). 

\s\, and [\ion{N}{i}] lines are also observed in the NIR part of the 
spectrum. They consist of four closely spaced lines for each species
 (\s\, at 1.029, 1.032, 1.034, and 1.037\um\, and [\ion{N}{i}] at 1.0400,
 1.0401, 1.04100, and 1.04104\um\,) having excitation energies 
higher than the optical lines ($\sim$35\,000 and 40\,000 K, respectively). 
Thus, in principle, the \s1.03$\mu$m 
lines can be combined with the optical \s\, lines to give an 
independent estimate of $T_{e}$. On the other hand, the [\ion{N}{i}] lines at 
1.04\um\, can be used with the [\ion{N}{ii}] optical lines to 
determine the ionization fraction. 

\subsection{Iron lines}

Most of the observed \fe\, lines originate from transitions among the
first 16 fine structure levels. As illustrated in Nisini et al. (2002),
electron densities in the range 10$^{3}$--10$^{5}$ cm$^{-3}$
can be derived from line ratios of IR lines, 
1.64/1.60$\mu$m and 1.64/1.53$\mu$m in particular, 
while the ratios 1.64/1.25$\mu$m and 1.64/1.32$\mu$m can be used 
to determine the visual extinction (see Appendix B). 
In the optical part of the spectrum, we also observe transitions coming 
from levels at higher excitation temperatures. Consequently, the ratios
 between selected optical and IR \fe\, lines are a good diagnostic tool for the measurement
of the electronic temperature in the range 5\,000--20\,000 K 
(Nisini et al. 2002, Pesenti et al. 2003). 

The combination of the IR
and optical spectrum of ionized Fe can therefore be used effectively to give
an independent diagnostic of temperature and density, which, relying
on ratios from lines of a single species, is not affected by any 
assumption of abundances. 
For our analysis, we use a statistical equilibrium model
which considers the first 16 \fe\, levels, the collisional strengths from 
Zhang \& Pradhan (1995)
and the transition probabilities from Nussbaumer \& Storey (1988).
An example of a diagnostic diagram for such an analysis is shown in Fig.
\ref{fig:fe_diag}. Similar diagrams are also presented in Nisini et al. (2002)
and Pesenti et al. (2003).

We have also
detected, in some of the knots, the lines at 7155, 7172, and 7453\AA\, which
connect the doublet $a^2\!G_{9/2,7/2}$ with the $a^4\!F$
levels. As shown in Bautista \& Pradhan (1998) and Hartigan et al. (2004),
the ratio between the 7155\AA\, and 8617\AA\, transitions is sensitive
to electron densities in the range $n_e \sim$10$^{5}$-10$^{7}$ cm$^{-3}$,
signalling the presence of high density conditions.

\subsection{Other atomic/ionic lines}

Other prominent atomic/ionic lines observed in the spectra include transitions
from the refractory species [\ion{C}{I}], [\ion{Ca}{II}],
[\ion{Cr}{II}] and [\ion{Ni}{II}]. 

The [\ion{C}{I}]$\lambda\lambda$9824,9850 lines 
can be excited both by collisions with neutrals and electrons in a partially
ionized medium, or by recombination (Escalante \& Victor 1990). In this latter case,
 other lines from the recombination cascade, such as
the 1.07 and the 1.17$\mu$m transitions, which are expected to be only a factor 
1-10 weaker than the 9840\AA\, lines, should also be detected (Walmsley et al. 2000). 
Their non-detection in any of our spectra would imply 
that the [\ion{C}{I}] lines are collisionally excited in the HH1 jet.
The presence of strong [\ion{C}{I}] lines also implies low excitation
conditions, and in fact the ratio [\ion{C}{I}]/H$\alpha$ is a very good
indicator of excitation in HH objects.

We detect [\ion{Ca}{II}] lines originating from transitions among the first 
five Ca$^+$ levels. In particular,
we detect the two forbidden lines at 7291 and 7324\AA\, connecting the $^2\!D_{5/2,3/2}$
levels with the ground state, as well as the permitted lines at 8542 and 8662\AA,
connecting the $^2\!P_{3/2,1/2}$ levels with the $^2\!D_{5/2,3/2}$.
These latter two transitions can be collisionally excited, being only at
25\,000 cm$^{-1}$ from the ground state or, alternatively, 
photoexcited through the  \ion{Ca}{II} K and H lines. Thus, the ratios 
among the observed lines can be 
used as a diagnostic to see if fluorescent excitation is important.

Ni$^+$ is expected to coexist with Fe$^+$, given their similar ionization
potential (7.9 and 7.6 eV, respectively). It has been found to be overabundant 
with respect to the solar value both in HH objects and in other gaseous nebulae
such as supernova remnants (Beck-Winchatz et al. 1996, Bautista et al. 1996).
In the latter, the observed strong [\ion{Ni}{ii}] emission has been explained by 
the contribution of fluorescent excitation in a low density environment. Indeed,
the ratio between the optical
lines, at 7412 and 7378\AA\, can be used as a discriminator between collisional and
fluorescent excitation (Bautista et al. 1996).

Finally, in the spectra of the brightest knots we also detected the 
two [\ion{P}{ii}] transitions at 1.148 and 1.189$\mu$m. Such lines have
excitation properties very similar to the \fe\, lines but P is a 
non-refractory element. This makes the IR ratio \fe/[\ion{P}{ii}]
a good tool to estimate Fe depletion (Oliva et al. 2001).

\section{Determination of $A_V$ from \fe\, lines}

\begin{figure}[!ht]
\resizebox{\hsize}{!}{\includegraphics{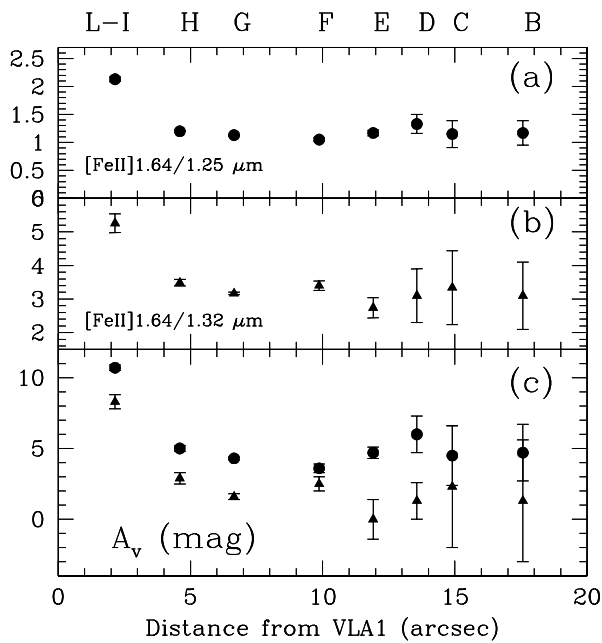}}
\caption{\label{fig:rat-av} Observed variations along the HH1 jet of the 
reddening sensitive \fe\, line ratios 1.64/1.25$\mu$m (a) and 1.64/1.32$\mu$m (b).
Panel (c) shows the derived $A_V$ values from the two ratios 
assuming an interpolation of the Rieke \& Lebofsky (1985) extinction law
and transition probabilities from Nussbaumer \& Storey (1988). Circles
and triangles are the $A_V$ values measured from the 1.64/1.25 and
1.64/1.32 line ratios, respectively.}
\end{figure}

Visual extinction can be estimated from pairs of lines
originating from the same upper level, providing the emission is
optically thin. In this case, their theoretical intensity ratio 
depends only on frequency and transition probabilities, and not on
 physical conditions in their emission region. 
There are several \fe\, lines which have this property, and 
the ratio \fe1.64/1.25$\mu$m (hereafter 1.64/1.25) is most commonly used 
for this purpose, as these lines are the brightest of the near IR \fe\, spectrum (Gredel et al. 1994,
Nisini et al. 2002). Among detected \fe\, lines, the bright transition
at 1.32$\mu$m also comes from the same upper level as the 1.64 and 1.25$\mu$m lines,
and thus the ratio \fe1.64/1.32$\mu$m (hereafter 1.64/1.32) gives an 
independent check of reddening.
In Fig. \ref{fig:rat-av} we plot the observed 1.64/1.25 and 
1.64/1.32 ratios in the different jet positions. This plot shows
the reddening variations along the jet irrespective of the absolute value
of the visual extinction, indicating that the reddening is almost constant
from knots A to H and sharply increases in knots L-I. This is consistent with
 Reipurth et al. (2000) who estimated that the visual extinction
in the inner portion of the jet is larger, by $\sim$ 4 mag, 
than that in the external knots. To convert the observed ratios to A$_V$ values,
we adopt the radiative transition probabilities given by Nussbaumer \& Storey
(1988), which imply intrinsic ratios 1.64/1.25=0.73 and 1.64/1.32=2.81. We
further use an interpolation of the near IR Rieke \& Lebofsky (1985) extinction law
of the form $A_{\lambda}/A_V=0.42\,\lambda(\mu m)^{-1.75}$ (Draine 1988).
In Figure \ref{fig:rat-av} we also plot the $A_V$ values derived from the two
 line ratios. Note that the extinction estimated
from the 1.64/1.25 ratio is always about a factor of two larger than the
values estimated from the 1.64/1.32 ratio. 

Such differences can be critical when estimating the intrinsic ratio 
of lines far apart in wavelength. Thus, we need to investigate the 
possible causes of such a discrepancy in more detail.

The observed discrepancy cannot originate from
any systematic observational effect. The three lines (1.64, 1.25, 
and 1.32$\mu$m) are all observed with the same grism and thus are not affected by
any problem of intercalibration. They are detected at large S/N, 
implying that any contamination due to OH sky line residuals
(which could affect the 1.64$\mu$m line) is negligible.
We then checked for observations of these three lines in other objects.
In Nisini et al. (2002) the $A_V$ values estimated from the 1.64/1.25 ratio
in a number of HH objects were always higher than the value 
derived from optical observations. Moreover we have verified that 
extinction values derived from the 1.64/1.32 ratio are more consistent 
with the optical determinations. The same effect, i.e. $A_V(1.64/1.25) > A_V(1.64/1.32)$,
can be found in various gaseous nebulae where these lines have been
detected with sufficent accuracy (e.g. Orion B, Walmsley et al.
2000, and supernova remnants, Oliva et al. 1990). 
 Finally, in the HH1 bow shock Noriega-Crespo \& Garnavich (1994)
estimated an Av = 6.7 using the 1.64/1.25 ratio, pointing out
the discrepancy between this value and that measured through
optical line ratios.

The most likely explanation for such a discrepancy is probably uncertainty
in the transition probabilities of the lines involved. More recent
calculations of such probabilities, by Quinet et al. (1997), show discrepancies
with respect to the Nussbaumer \& Storey (1988) values as large as 20\%. 
With the Quinet et al.(197)  $A$ values, the 
theoretically expected ratios become 1.64/1.25=0.965 and 1.64/1.32=3.67
\footnote{In a note added in proof, Quinet et al. (1995) evaluated the limitations
of their $A_{ij}$ calculations done considering only three \fe\, electronic 
configurations. While in general
the inclusion of the additional 3d$^{5}$4d$^2$ configuration does not change
the radiative rates by more than few percent, for the explicit case 
of the a$^4$F--a$^4$D transitions (which includes the 1.64$\mu$m transition)
the correction is about 40\%. If real, this would cause the 1.64/1.25 theoretical
ratio to become again very similar to the 1.64/1.25 NS88 expected ratio.}.
If applied to our data, such ratios produce on average $A_V$ smaller
than the values derived with the NS88 data, but the discrepancy between
the 1.64/1.25 and 1.64/1.32  estimated values remains. In knot G, for
example, we derive, with the Quinet et al. (1997) values, 
$A_V(1.64/1.25)$= 1.6 mag, but $A_V(1.64/1.32)< 0$. In general, the $A_V(1.64/1.32)$ value estimated
with the parameters by Quinet et al. (1997) tends to be always negative, suggesting that
the theoretical value is probably overestimated.

Finally, we point out that the higher $A_V$ values
derived from the 1.64/1.25(NS88) ratio, turn out to be inconsistent with
the ratios of other observed transitions irrespective of the assumed
physical conditions. For example the \fe1.64/0.862 ratio in 
 knot G, if corrected for an $A_V$ larger than 4, is outside
the range of expected theoretical ratios for any temperature (see Fig. 4).
The same applies to the ratio [\ion{S}{ii}]0.67/1.03 (see also
the discussion in Section 4.2). 
In view of the above discussion, we suggest that the 1.64/1.32 ratio from 
NS88 gives, for the moment, the most reliable determination of the $A_V$ value.

\begin{acknowledgements}
We thank Silvia Medves for letting us use her unpublished \s\, data on the HH1 jet.
We are also grateful to Paola Caselli and Malcolm Walmsley for useful discussions on
grain disruption processes. This work was partially supported by the European 
Community's Marie Curie Research and Training Network JETSET (Jet Simulations, 
Experiments and Theory) under contract MRTN-CT-2004-005592.
\end{acknowledgements}


\begin{thebibliography}{}
\bibitem{} Asplund, M., Grevesse, N. \& Sauval, A.J., 2004, in Cosmic Abundances 
as Records of Stellar Evolution and Nucleosynthesis, ASP Conference Series, 
Vol. XXX (astro-ph/0410214)
\bibitem{} Bacciotti, F. \& Eisl\"offel, J., 1999, \aap, 342, 717 (BE99)
\bibitem{} Bacciotti, F. 2002, in ''Emission Lines from Jet Flows'', RMxAA,
Vol. 13, p. 8
\bibitem{} Bally, J., Heathcote, S., Reipurth, B., Morse, J., 
Hartigan, P. \& Schwartz, R. 2002, AJ, 123, 2627
\bibitem{} Bautista, M.A., Peng, J., \& Pradhan, A. K. 1996, ApJ, 460, 372
\bibitem{} Bautista, M.A. \& Pradhan, A. K. 1998, ApJ, 492, 650 
\bibitem{} Beck-Winchatz, B., B\"{o}hm, K.H. \& Noriega-Crespo, A., 1996,  AJ, 111, 346
\bibitem{} B\"{o}hm, K.H. \& Matt, S., 2001, PASP, 113, 158
\bibitem{} Chidichimo, M.C. 1981, J. Phys. B 14, 4149
\bibitem{} Correia, J.C., Griffin, M., Saraceno, P. 1997,A\&A, 322, L25
\bibitem{} Davis, C.J., Eisl\"offel, J., Ray, T.P. 1994, ApJ 426, L93
\bibitem{} Davis, C.J., Smith, M. D. \& Eisl\"offel, J. 2000, MNRAS, 318, 747
\bibitem{} Davis, C.J., Hodapp, K.W. \& Desroches, L. 2001, A\&A, 377, 285
\bibitem{} Davis, C.J., Whelan, E., Ray, T. P. \& Chrysostomou, A. 2003, A\&A, 
397, 693
\bibitem{} Draine, B.T. 1989, in ``Infrared Spectroscopy in Astronomy'', ESA-SP290,
p. 93
\bibitem{} Draine, B.T. 2003, in ``The Cold Universe:Saas-Fee Advanced Course 32'',

astro-ph/0304488
\bibitem{} Dougados, C., Cabrit, S., Lavalley, C. \& M\'{e}nard, F. 2000,
A\&A, 357, 61
\bibitem{} Eisl\"offel, J., Mundt, R., B\"ohm, K.-H. 1994, AJ 108, 1042
\bibitem{} Eisl\"offel, J., Smith, M.D., Davis, C.J., 2000, A\&A 359, 1147
\bibitem{} Escalante, V. \& Victor, G.A. 1990, ApJSS, 73, 513
\bibitem{} Giannini, T., Nisini, B. \& Lorenzetti, D. 2001, ApJ, 555, 40
\bibitem{} Giannini, T., Nisini, B., Caratti o Garatti, A. \& Lorenzetti, D., 2002, ApJL, 570, 33
\bibitem{} Giannini, T., McCoey, C., Caratti o Garatti, A., Nisini B., Lorenzetti
D. \& Flower, D.R. 2004, A\&A, 419, 999
\bibitem{} Gredel, R. \& Reipurth, B., 1994, \aap, 289, L19
\bibitem{} Grevesse, N. \& Sauval, A.J. 1998, Space Sci. Rev. 85, 161
\bibitem{} Jones, A.P. 2000, \jgr, 105, 10257
\bibitem{} Hartigan, P., Raymond, J., Hartmann, L. 1987, ApJ, 316, 323
\bibitem{} Hartigan, P., Morse, J.A. \& Raymond, J. 1994, ApJ, 444, 943
\bibitem{} Hartigan, P., Edwards, S. \& Pierson, R. 2004, ApJ, 609, 261
\bibitem{} Hester, J.J., Stapelfeldt, K.R. \& Scowen, P.A. 1998, AJ, 116, 372
\bibitem{} Hollenbach, D. 1985, Icarus, 61, 36
\bibitem{} Landini, M. \& Monsignori Fossi, B.C. 1990, A\&AS, 82, 229
\bibitem{} Medves S. 2003, Graduation thesis, Pisa University
\bibitem{} Mendoza, C. 1983, in Planetary Nebulae, ed. D. R. Flower (Dordrecht: Reidel), 
IAU Symp., 103, 143
\bibitem{} Molinari, S. \& Noriega-Crespo, A., 2002, AJ, 123, 2010
\bibitem{} Moro-Mart\'{\i}n, A., Cernicharo, J., Noriega-Crespo, A., Mart\'{\i}n-Pintado,
J. 1999, ApJ, 520, L111
\bibitem{} Mouri, H. \& Taniguchi, Y., 2000, ApJL, 534, L63
\bibitem{} Nisini, B., Caratti o Garatti, A., Giannini, T., \& Lorenzetti, 
D. 2002, A\&A, 393, 1035
\bibitem{} Noriega-Crespo, A. \& Garnavich, P.M. 1994, Rev. Mex. A.A., 28, 173
\bibitem{} Noriega-Crespo, A., Garnavich, P.M., Curiel, S., 
Raga, A. \& Ayala, S. 1997, ApJL, 486, L55
\bibitem{} Nussbaumer, H. \& Storey, P.J. 1988, A\&A, 193, 327
\bibitem{} Oliva, E., Moorwood, A.F.M. \& Danziger, I.J. 1990, A\&A, 240, 453
\bibitem{} Oliva, E., Marconi, A., Maiolino, R. et al. 2001, A\&A, 369, L50
\bibitem{} Quinet, P., Le Dourneuf, M., \& Zeippen, C. J. 1996, 
A\&AS, 120, 361 
\bibitem{} Reipurth, B., Heathcote, S., Yu, K.C., Bally, J. \& Rodr\'{\i}guez, L.F.,
2000, ApJ, 534, 317
\bibitem{} Rieke, G. H., \& Lebofsky, M. J. 1985, ApJ, 288, 618
\bibitem{} Rodr\'{\i}guez, L.F., Delgado-Arellano, V.G., G\'{o}mez Y. et al. 2000,
AJ, 119, 882
\bibitem{} Solf, J., Bo\"hm, K.H. \& Raga, A.C. 1988, ApJ, 334, 229
\bibitem{} Solf, J., Raga, A.C. , Bo\"hm, K.H. \& Noriega-Crespo, A. 1991, AJ,
102,1147
\bibitem{} Stancil, P.C., Havener, C.C., Krstic, P.S. et al. 1998, ApJ, 502, 1006
\bibitem{} Strom, S.E., Strom, K.M., Grasdalen, G.L., Capps, R.W. \& 
Thompson, D. 1985, AJ, 90, 2575
\bibitem{} Walmsley C.M., Natta, A., Oliva, E., Testi, L. 2000, A\&A, 364, 301
\bibitem{} Zhang, H.L. \& Pradhan, A.K. 1995, A\&A, 293, 953
\end{thebibliography}
\end{document}